\documentclass[12pt]{article}
\usepackage{amsmath, setspace,geometry,color}
\usepackage{graphicx}
\usepackage{lineno}
\usepackage{tabularx}

 \makeatletter       
 \renewcommand{\@biblabel}[1]{#1.}
\newcommand{\beginsupplement}{%
        \setcounter{table}{0}
        \renewcommand{\thetable}{S\arabic{table}}%
        \setcounter{figure}{0}
        \renewcommand{\thefigure}{S\arabic{figure}}%
     } 
 
 \makeatother
\begin{document}
\doublespacing
\noindent
{\LARGE\textbf{New classification method of volcanic ash samples using statistically determined grain types}}
\\
\\
{\large R. Noguchi$^{1*}$, H. Hino$^{2}$, N. Geshi$^{3}$, S. Otsuki$^{3}$, and K. Kurita$^{4}$}\\
\\
\doublespacing
1. \textit{Volcanic Fluid Research Center, School of Science, Tokyo Institute of Technology, 2-12-1, Ookayama, Meguro-ku, Tokyo 152-8551, Japan}\\
2. \textit{Department of Computer Science, University of Tsukuba, 1-1-1 Tenoudai, Tsukuba, Ibaraki 305-8573, Japan}\\
3. \textit{Geological Survey of Japan, AIST.  AIST Build. No. 7, 1-1-1 Higashi, Tsukuba, Ibaraki 305-8567, Japan}\\
4. \textit{Earthquake Research Institute, the University of Tokyo, 1-1-1 Yayoi, Bunkyo-ku, Tokyo, 113-0032, Japan}\\

$*$ Corresponding author: r-noguchi@ksvo.titech.ac.jp
\newpage
 \linenumbers*[1]
\section*{abstract}
\textbf{We developed a method to classify volcanic ash samples by introducing statistically determined grain types.
Using more than 10,000 numbers of automatically measured grain data (parameters of grain shape and transparency) and the cluster analysis, we made grain types without human eyes.
By components of the grain type in each samples, we classified samples from types of basaltic monogenetic volcanoes: 1) Funabara scoria cone, Izu Peninsula, Japan (magmatic eruption origin); 2) Nippana tuff ring, Miyakejima, Japan (phreatomagmatic eruption origin); and 3) rootless cones in Myvatn, Iceland (rootless eruption origin).
We tested two cases; using grain shape parameters only, and both of grain shape parameters and transparency values.
It is found that the sample classification is more consistent with their eruption style in the case of using both parameters of grain shape and transparency.
By sampling several layers of an outcrop, this procedure can be used to interpret changes in eruption/fragmentation style during a volcanic event.
Furthermore, this procedure might be applicable to other aims such as sedimentology and planetary science.}

\section{Introduction}\label{Introduction}

The grain morphology of volcanic pyroclasts provides important information that allows us to infer eruption styles and mechanisms from microscopic images of volcanic ash.  
Grain analysis can reveal the characteristics of magma vesiculation and subsequent fragmentation (e.g., \cite{Heiken1985}).
Several studies have attempted to parameterize grain shapes to explore possible relationships with their formation and fragmentation processes (\cite{Dellino1996}; \cite{Dellino2001}; \cite{Maria2002}; \cite{Maria2007}; \cite{Liu2015}; \cite{Rausch2015}; \cite{Schmith2017}).

Owing to recent developments of measurement instruments, we can easily parameterize the visual characteristics (shape and luminance) of thousands of grains in a short time, and can perform quantitative analysis for large numbers of volcanic ash grains.
Introducing an automated particle analyzer (APA), \cite{Leibrandt2015} tested several measurement procedures, before presenting an efficient measurement protocol for volcanic ash.
In their system, the operating duration is 35 minutes for 5000 ash grains, and several shape parameters are measured for each ash grains.

When analyzing this type of multivariate data, the parameter selection step is essential. 
With this in mind, \cite{Liu2015} used a cluster analysis to determine four optimal parameters (solidity, convexity, axial ratio, and form factor), which can effectively account for the morphological variance of grains.
\cite{Maria2002,Maria2007} applied the fractal spectrum technique to volcanic ash grains, and used the fractal values (they called as variables) to perform 1) cluster analyses for ash grains in each sample (consists of 20 ash grains), and 2) principal component analyses for 140 of ash grains.
Thus, we can use a combination of grain shape parameters and statistical techniques to determine eruption styles and characteristics from the analysis of volcanic ash.
Although this approach has been shown to be effective, it remains necessary to verify the applicability of this procedure to wider data sets.
Specifically, verification should involve simple eruption episodes, such as monogenetic eruptions.
Furthermore, in comparison among samples, characteristics of magma such as chemical composition and phenocryst content should be considered.

When analyzing volcanic ash samples, there is a major problem; how to compare among samples.
Field studies entail collecting many samples from several layers, outcrops, volcanic edifices, and volcanic systems.
Sieving these samples to microscopic sizes involves several thousand ash grains.
Many previous volcanic ash studies have focused on ash classification in one or limited numbers of samples.
For sample classification, most previous studies have used one value for one parameter in one sample (e.g., \cite{Dellino2001}); it far from traditional component proportion analysis under microscopic observation by human eyes, and made difficulty to its interpretation.
In this background, we have to discriminate both of volcanic ash grains and ash samples which comparable to microscopic observations.

In this study, we adopt cluster analyses for automatically-measured large volcanic ash datasets.
Firstly we made "grain types" by quantitative grain parameters without human eyes, then classified samples using its proportion of each grain types.
We construct a statistical analysis procedure for volcanic ash samples aimed 1) to compare samples from several layers, outcrops, volcanic edifices, and volcanic systems, and 2) to quantify the fragmentation degree and the effect of external water to volcanic explosions.

\section{Methodology}
\subsection{Prepared samples}

We use ash samples from non-altered monogenetic volcanoes to simplify the analysis.
We collected 18 ash samples from three locations in Japan and Iceland (Table \ref{table_list}, Fig.S1,S2), which were formed during three different types of monogenetic eruption: magmatic (vaporization of volatiles in magma), phreatomagmatic (magma-water interaction), and rootless eruptions (explosive lava-water interaction).
Since the chemical compositions and phenocryst characteristics (composition, mode, and size) of host magma have resemblances (Table \ref{Petrological}), we selected ash samples from these three locations, and assumed that their phenocryst and magmatic features were the same.
Each sample is composed of many ash grains, and the grain characteristics such as shape and size differ for each sample.
The grain size distribution for each sample is shown in Fig.S3.

\newpage
\begin{table}[htbp]
\centering
\caption{List of samples used in this study. DRC: double rootless cone, SRC: single rootless cone.}
\small
\begin{tabular}{|c|c|c|c| }
 \hline
Eruption type (sample location) & Sample ID & Number of grains & Note\\
\hline
Magmatic   & FN15101201 & 131 & Lower layer \\
 (Funabara, Izu Peninsula)& FN15101205 & 262 & Upper than 01\\
Japan& FN15101206 & 206 & Upper than 05\\
& FN15101207 & 87 & Upper than 06\\
& FN15101208 & 168 & Upper than 07\\
\hline
Phreatomagmatic & NP15113001 & 1851 & Lower layer\\
 (Nippana, Miyakejima) & NP15113002 & 707 & Upper than 01\\
Japan& NP15113003 & 428 & Upper than 02\\
& NP15113004 & 1125 & Upper than 03\\
& NP15113005 & 708 & Upper than 04\\
& NP15113006 & 796 & Upper than 05\\
& NP16102407 & 863 & Upper than 06\\
\hline 
Rootless  & MY13091004 & 923 & DRC outer, middle layer\\
 (Myvatn, N Iceland)& MY13091006 & 1065 & DRC inner, lower layer\\
Iceland& MY13091305 & 686 & SRC, lower layer\\
& MY13091306 & 670 & SRC, middle layer\\
& MY13091402 & 1479 & SRC, upper layer, collected in Hagi\\
& MY13092002 & 965 & SRC, lower layer\\
 \hline
\end{tabular}
\label{table_list}
\end{table}

\newpage
\begin{table}[htbp]
\begin{center}
\caption{Petrological information of samples used in this study.}
\small
\begin{tabular}{|p{3cm}p{1cm}|p{4cm}|p{4cm}|p{4cm}|llll}
\hline
 &  & Magmatic & Phreatomagmatic & Rootless\\
 &  & Funabara & Nippana (Miyakejima 1983) & Myvatn (Younger Laxa lava)\\
\hline
Bulk composition & SiO$_{2}$ & 51.23 & 52.91 & 49.3\\
 & TiO$_{2}$ & 1.09 & 1.42 & 1.1\\
 & Al$_{2}$O$_{3}$ & 17.3 & 14.98 & 14.98\\
 & Fe$_{2}$O$_{3}$ & 3.47 & 14.33 & 0.9\\
 & FeO & 6.71 & N/A & 9.27\\
 & MnO & 0.17 & 0.24 & 0.19\\
 & MgO & 6.85 & 4.13 & 6.96\\
 & CaO & 9.92 & 9.11 & 12.63\\
 & Na$_{2}$O & 2.71 & 2.67 & 2.15\\
 & K$_{2}$O & 0.41 & 0.53 & 0.91\\
 & H$_{2}$O+ & 0.4 & N/A & 0.79\\
 & H$_{2}$O-- & 0.02 & N/A & 0.00\\
 & P$_{2}$O$_{5}$ & 0.28 & 0.14 & 0.28\\
 & Total & 100.56 & 100.46 & 99.46\\
 \hline
 Phenocrysts &  & olivine, pyroxene (augite and hypersthene), quartz & plagioclase, clinopyroxene, magnetite & plagioclase, pyroxene, olivine\\
\hline
Mode of phenocrysts [\%] &  & 3.7\% & 5--8\%: plagioclase, $<$0.1--0.3\%: clinopyroxene, $<$0.2\%: magnetite & less than 10\%\\
\hline
Phenocrysts size &  & olivine: 1 mm, hypersthene: $<$1 mm, quartz: 3 mm (occasionally), magnetite: 0.2 mm & plagioclase: 0.6--0.8 mm (major axes), 0.2--0.3 mm (minor axes), clinopyroxene: 0.2--0.4 mm, magnetite: 0.2 mm & plagioclase: 0.2--0.6 cm (1.5--2 cm for large)\\
 \hline
References &  & \cite{Yusa1970}; \cite{Hamuro1985} & \cite{Fujii1984}; \cite{Aramaki1986} (Sample ID: MYK-11) & \cite{Thorarinsson1953}; \cite{Hoskuldsson2010}\\
 \hline
\end{tabular}
\label{Petrological}
\end{center}
\end{table}

\subsubsection{Funabara scoria cone}\label{Funabara scoria cone}
Funabara scoria cone in Izu Peninsula (Fig.S1A) belongs to the Higashi-Izu monogenetic volcano group (HIMVG; \cite{Koyama1991}).
The monogenetic volcanic activity in HIMVG began in 0.26$\pm$0.02 Ma (\cite{Hasebe2001}).
Today, there are over 60 monogenetic volcanoes in this area.
This volcanic field displays a variety of magmatic compositions: basaltic, andesitic, dacitic, and rhyolitic (e.g., \cite{Hamuro1985}).
The age of the Funabara scoria cone was estimated as 0.20$\pm$0.08 -- 0.22$\pm$0.09 Ma by \cite{Hasebe2001}.
The rock type is basalt with 50.91--51.23 wt\% of SiO$_2$, and the mode of phenocrysts is 3.7\% (\cite{Hamuro1985}).
Currently, the Funabara scoria cone is being quarried, revealing a clear stratigraphic profile (Fig.\ref{Outcrops.jpg}A).
At the outcrop, we took five samples from bottom to top.

In 2$\phi$--3$\phi$ scale, Funabara samples are dominant red-oxidized opaque grains (Fig.\ref{microscopic_image.jpg}).
Under the incident lighting, they show glassy surface.
They contains microlites (plagioclase, olivine, and magnetite) in microscopic observation ($\times$ 500 magnification).
Some of them are coated with red-oxidized magna, therefore they are opaque in the transmitted lighting.
Funabara samples also contain transparent grains: brownish yellow grains and free-crystals.
Brownish yellow grains show glassy surface, and are more dominant in FN15101206.
Free-crystals are plagioclase, olivine, and pyroxene.

\subsubsection{Nippana tuff ring}\label{Nippana tuff ring}
\begin{figure}[htbp]
\begin{center}
\includegraphics[width=15cm]{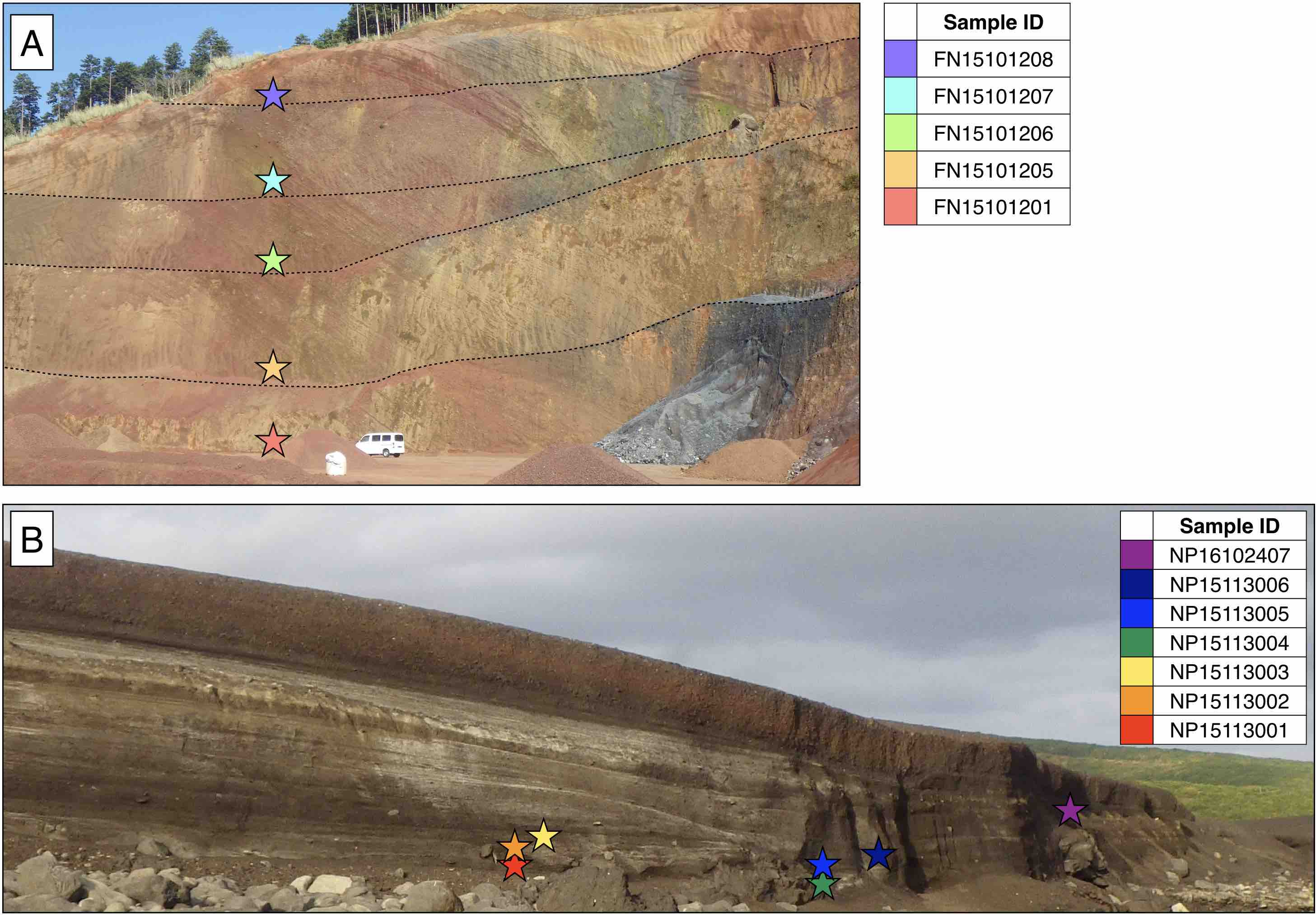}
\caption{Sampling outcrops of Funabara scoria cone (A) and Nippana tuff ring (B).}
\label{Outcrops.jpg}
\end{center}
\end{figure}

Nippana tuff ring is a half-collapsed tuff ring in the south of Miyakejima Island, Japan (Fig.S1B).
On October 3, 1983, explosive interactions of magma and sea water formed a tuff ring due to the eruption center reaching a littoral area on the southwestern flank of the main edifice of Miyakejima (Oyama) through a fissure vent system (e.g., \cite{Sumita1985}; \cite{Aramaki1986}).
The rock type is augite basalt with 52.3--54.6 wt\% of SiO$_2$. and the mode of phenocrysts is less than 9\% (\cite{Aramaki1986}).
After formation, the edifice was half destroyed by subsequent typhoons and erosion (\cite{Sumita1985}, Fig.\ref{Outcrops.jpg}B).
We collected seven samples at the outcrop from lower to upper layers.

Nippana samples are rich in transparent grains (Fig.\ref{microscopic_image.jpg}).
Most of them are glass fragment which often contain microlites (plagioclase and magnetite) in microscopic scale ($\times$ 500 magnification for 2.5$\phi$--3$\phi$ grains).
The others are free-crystals (plagioclase).
NP16102407 contains black opaque glassy grains.
Qualitatively, characteristics of grain shapes are different among samples; inwardly convex shape (NP15113001, NP15113002, and NP15113003), and rectilinear edge (NP15113004, NP15113005, and NP15113006).

\subsubsection{Myvatn rootless cones}\label{Myvatn}
\begin{figure}[htbp]
\begin{center}
\includegraphics[width=15cm]{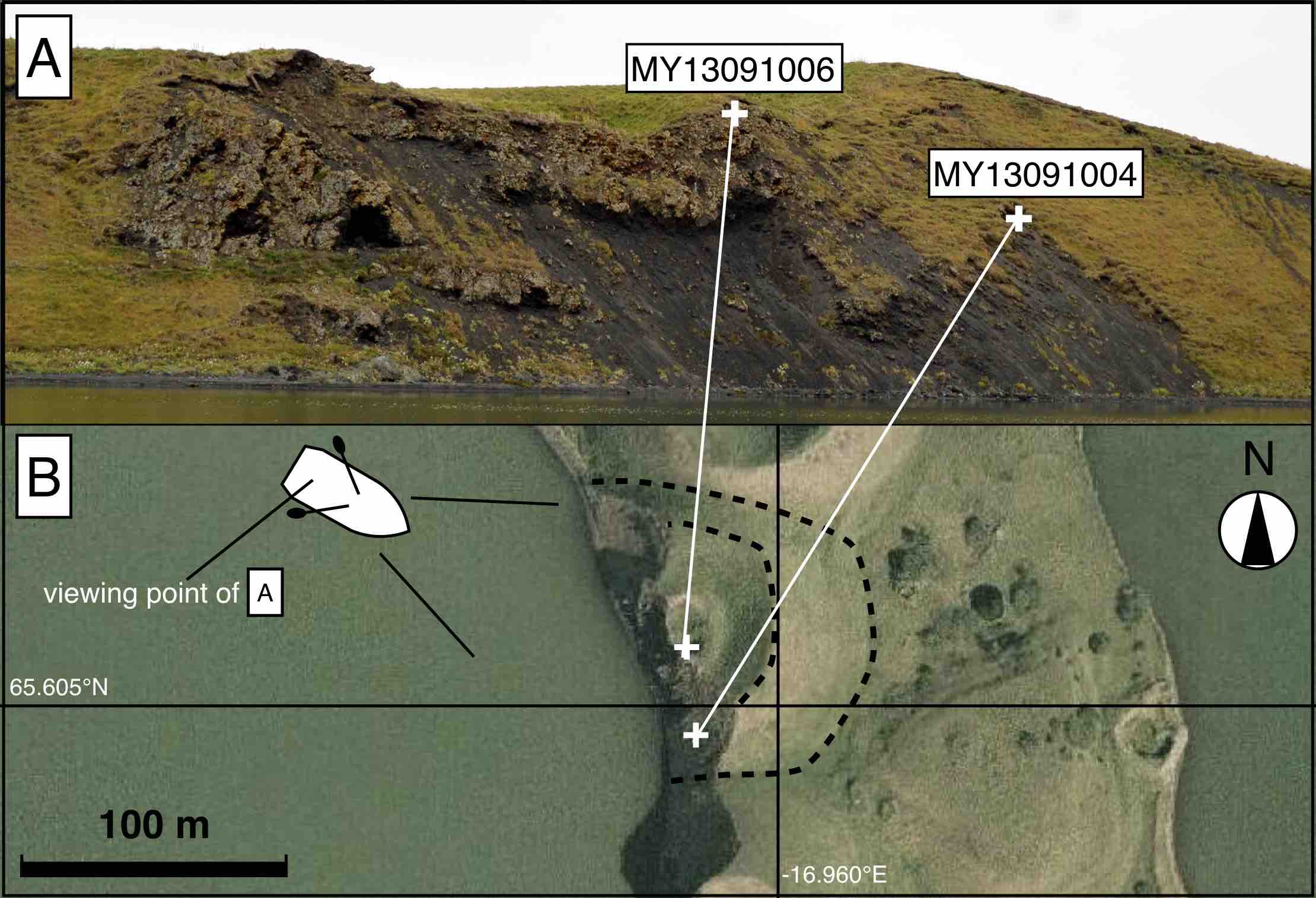}
\caption{Sampling outcrops of Myvatn rootless cones for MY13091004 and MY13091006.  This is a profile of double rootless cones which is thought to be formed by two stages of rootless eruptions (\cite{Noguchi2016})}
\label{MY130910.jpg}
\end{center}
\end{figure}

Myvatn is located in the north east of Iceland (Fig.S2).
The volcanic activity around this area is part of the \textit{Krafla Volcanic System}.
In this volcanic system, most effusive lava is basaltic, with 49\% wt\% of SiO$_2$ (\cite{Nicholson1990}).
The mode of phenocrysts is less than 10\% (\cite{Hoskuldsson2010}).
In 2170$\pm$38 cal yr BP (\cite{Hauptfleisch2012}), lava (\textit{Younger Lax\'a lava}) erupted from fissure swarms 12 km long on the east side of the lake.
The lava flowed over wetlands and the lake (old Lake Myvatn), and hundreds of rootless cones were formed by explosive interactions between the lava and lacustrine sediments (\cite{Thorarinsson1979}; \cite{Einarsson1982}).
Afterwards, the lava flowed down the Lax\'ardalur river, forming rootless cones, before reaching a northern bay.
In \cite{Hoskuldsson2010}, the duration of this eruption is estimated to have been 30 days from an analysis of the lava thickness based on an equation in \cite{Hon1994}.
Based on the microcrystalline growth data of olivine, \cite{Dolvik2007b} gives the timing of the rootless eruption as 1--2 days after the start of this eruption.
Some rootless cones have smaller cones inside of the summit crater (double rootless cones, DRCs, Fig.\ref{MY130910.jpg}; \cite{Noguchi2016}), though typical ones have simple conical edifice (single rootless cones, SRCs).
We collected samples from rootless cones in several areas of Myvatn (Fig.S2).
MY13091004--MY13091006 and MY13091305--MY13091306 were taken from same cone, respectively (Fig.\ref{MY130910.jpg}, S4).
MY13091402 was sampled in Hagi, 45 km distant from the fissure vents.

\begin{figure}[htbp]
\begin{center}
\includegraphics[width=15cm]{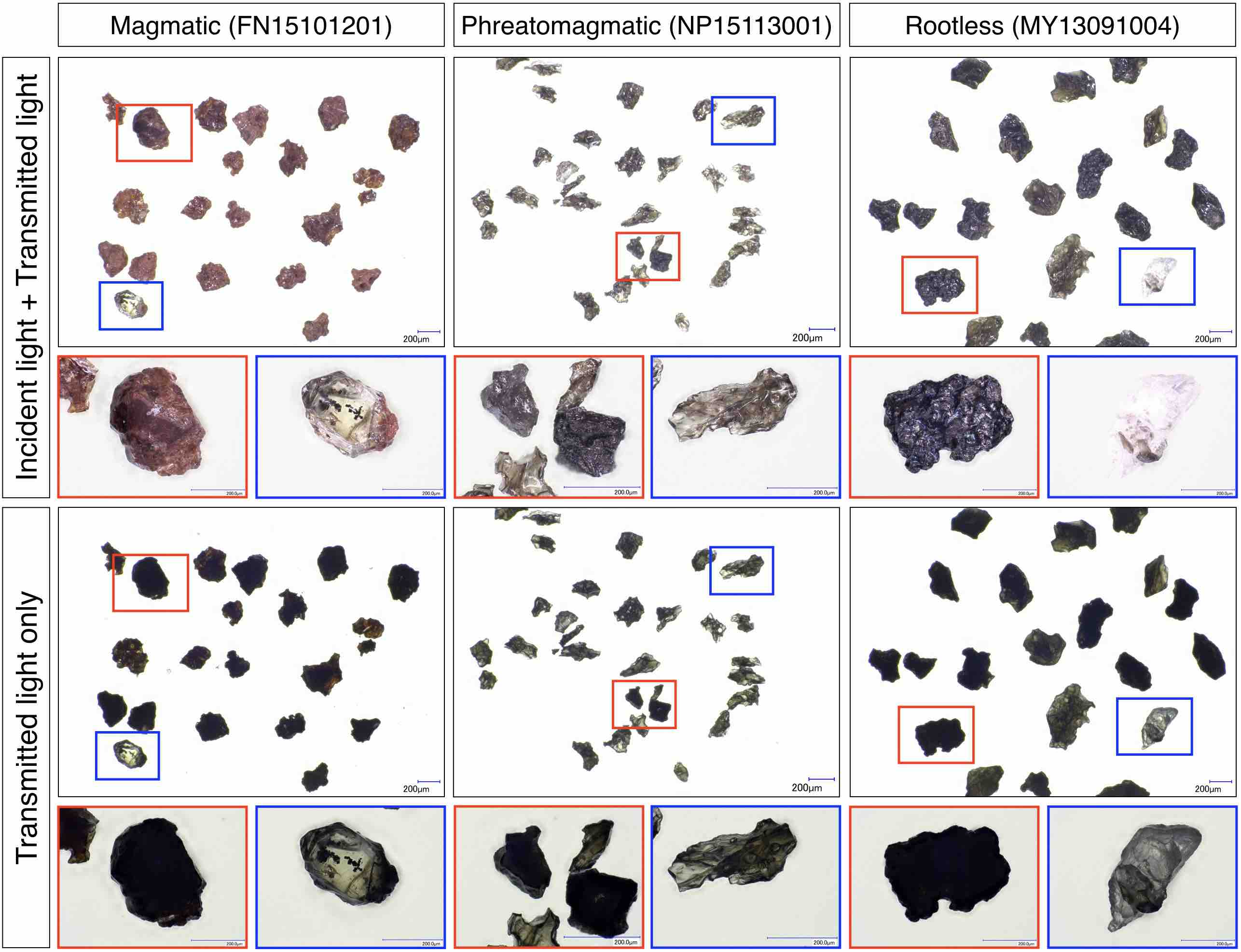}
\caption{Examples of microscopic pictures of volcanic ash samples in this study.  Red and blue boxes show example opaque and transparent grains, respectively.  In the setting of the incident light + the transmitted light, the transmitted light is used to reduce shadow of grains.  These microscopic pictures were taken by VHX-2000, KEYENCE at AIST.}
\label{microscopic_image.jpg}
\end{center}
\end{figure}

Most of grains in Myvatn samples are glassy (Fig.\ref{microscopic_image.jpg}).
In transparent glassy grains, there found microlites (plagioclase).
Plagioclases are also contained as free-crystals.
There exist dappled grains.
In the transmitted lighting, transparent grains are dominant in MY13091004 and MY13091305, and opaque grains are rich in MY13091306 and MY13091402, qualitatively.

\subsection{Grain shape and transparency measurement}

The grain shape parameters of the pyroclasts were measured using an APA: Morphologi G3S\texttrademark (Malvern Instrument\texttrademark) at the Geological Survey of Japan, AIST.
The detailed measurement method is shown in \cite{Leibrandt2015}.
This study measured sieved volcanic ash grains (2.5$\phi$ to 3$\phi$ fraction (160$\mu$m to 250$\mu$m) for Nippana samples (except NP16102407), and 2$\phi$--3$\phi$ (125--250 $\mu$m) for the others) at $\times$5 magnification.
At this magnification, the measurable grain diameter ranges from 6.5 to 420 $\mu$m; therefore, our sample sizes are appropriate for these measurement conditions.
To place the volcanic ash grains on the glass plate, we used a Sample Dispersion Unit (SDU) with 1.5 bar of injection pressure and a 20 ms injection time.
During the measurement, the illumination was set to diascopic (bottom light), under automatic light calibration (calibration intensity of 80.00 and intensity tolerance of 0.20).
The threshold for background separation (0--255) was set at 80 to obtain a sharp focus.
The measurement lasted approximately 40 min for each sample.
After the measurement, we excluded overexposed, unseparated, and cut-off grains.
To remove unwanted material (such as dust), we picked out grains with a solidity (a parameter of grain shape, $S_d$, sensitive to morphological roughness; \cite{Liu2015}) lower than 0.6.
In total, we collected parameterized data of 13,120 volcanic ash grains from the 18 samples.

\subsection{Grain parameters}

We used two-dimensional (2-D) projected images of ash grains to describe particle characteristics.
Using Morphologi, we can obtain seven grain shape parameters (circularity, high sensitive circularity, convexity, solidity, aspect ratio, and elongation) and two transparency values (intensity mean and intensity standard deviation).
\cite{Liu2015} described four shape parameters: convexity, solidity, axial ratio, and a form factor (HS circularity in Morphologi), which they adopted for grain shape analyses of volcanic ash.
In this study, we used the aspect ratio instead of the axial ratio, as it is not provided by Morphologi.

As well as grain shape, Morphologi can also measure luminance, which, under bottom lighting conditions, indicates grain transparency.
Information regarding the grain transparency of volcanic ash is important for identifying the glass and crystal components; however, most previous research has focused on grain shape alone, with the exception of \cite{Miwa2015}.

\begin{figure}[htbp]
\begin{center}
\includegraphics[width=15cm]{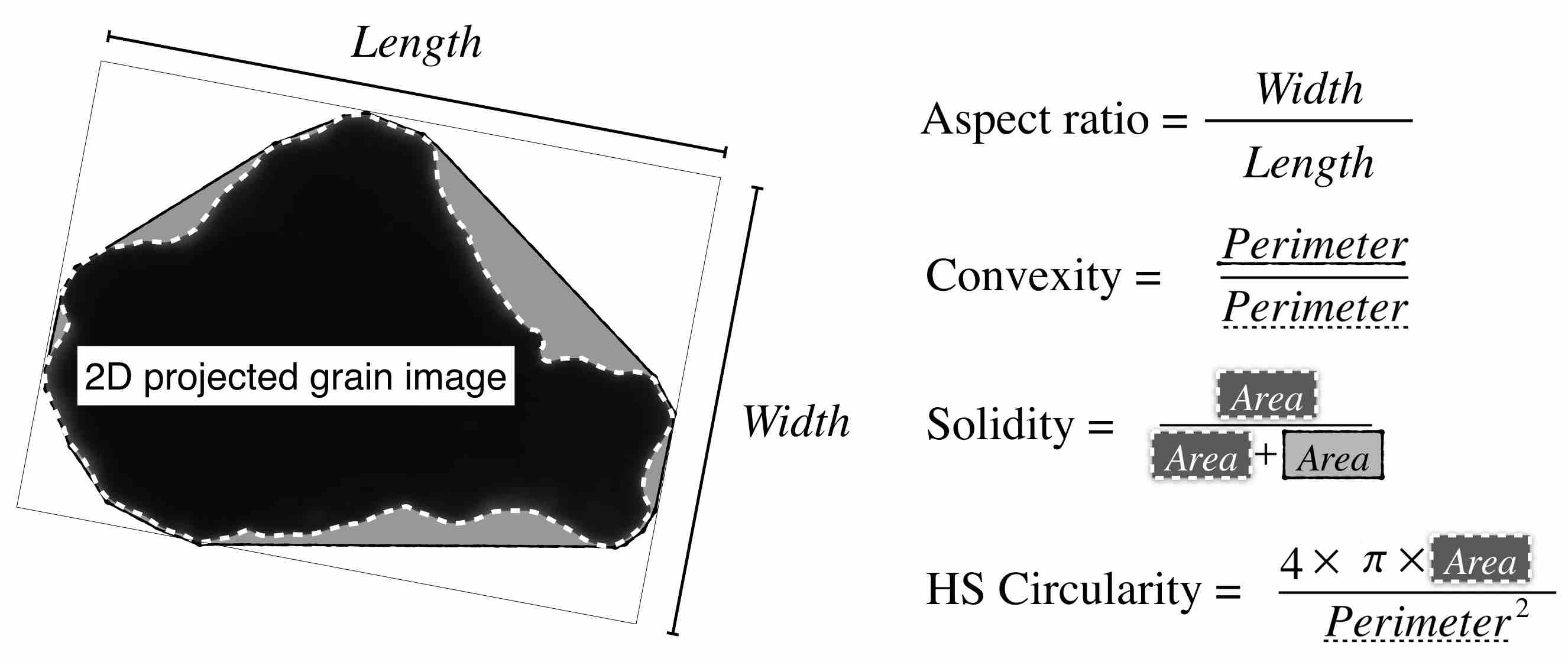}
\caption{Derivations of grain shape parameters used in this study.}
\label{GS_parameters.jpg}
\end{center}
\end{figure}

In this analysis, we chose six parameters: aspect ratio ($A_r$), convexity ($C_v$), solidity ($S_d$), HS circularity ($H_c$), intensity mean ($I_m$), and the standard deviation of the intensity ($I_{sd}$).
Derivations of these parameters are given by

\[
A_r=\frac{W}{L}
\]

\[
C_v=\frac{P_c}{P_g}
\]

\[
S_d=\frac{A_g}{A_g + A_c}
\]

\[
H_c=\frac{4 \times \pi \times A_g}{{P_g}^2}
\]

\[
I_m=\frac{\displaystyle\sum_{i=1}^{i=N} I_i}{N}
\]

\[
I_{SD}=\sqrt{\frac{{\displaystyle\sum_{i=1}^{i=N} {I_i}^2}-\frac{\left(\displaystyle\sum_{i=1}^{i=N} I_i\right)^2}{\displaystyle N}}{N}}
\]

where $W$ is length along the minor axis of the grain, $L$ is length along the major axis of the grain, $P_c$ is perimeter of the convex hull, $P_g$ is the perimeter of the grain, $A_c$ is area of the convex hull, $A_g$ is the area of the grain, $I_i$ is the intensity values (0--255) of pixel ($i$), and $N$ is the total number of pixels in the grains (\cite{Malvern2013}).
These parameters were calculated for each ash grain.
Definitions for each morphometrical parameter are shown in Fig.\ref{GS_parameters.jpg}.

\subsection{Data analysis}

A standard cluster analysis technique (\cite{Anderberg2014}) was applied to objectively and quantitatively classify samples.
The cluster analysis is one of the multivariate analysis method for unsupervised classification.
This statistical method is used in variety of fields such as marketing research and machine learning.
Using the cluster analysis, it is able to categorize data quantitatively (based on the cluster distance such as the Euclidean distance) and visually (as a dendrogram, a kind of tree diagram).
In this study, we adopted the hierarchical clustering method known as Ward's method (\cite{Anderberg2014}), because of its wide use and the ease of clustering result interpretation.
To apply the cluster analysis, we used standard machine learning and statistical methods to compare samples, namely, we represented the grains and samples by vectors composed of small number of features.
Feature vector representation of samples is convenient because we can use conventional methods of measuring the difference between samples, i.e., the Euclidean distance between feature vectors of different samples.
The cluster analysis was performed using the {\tt hclust} function equipped with the statistical computing environment R (\cite{RCoreTeam2016}).

\begin{figure}[htbp]
\begin{center}
\includegraphics[width=10cm]{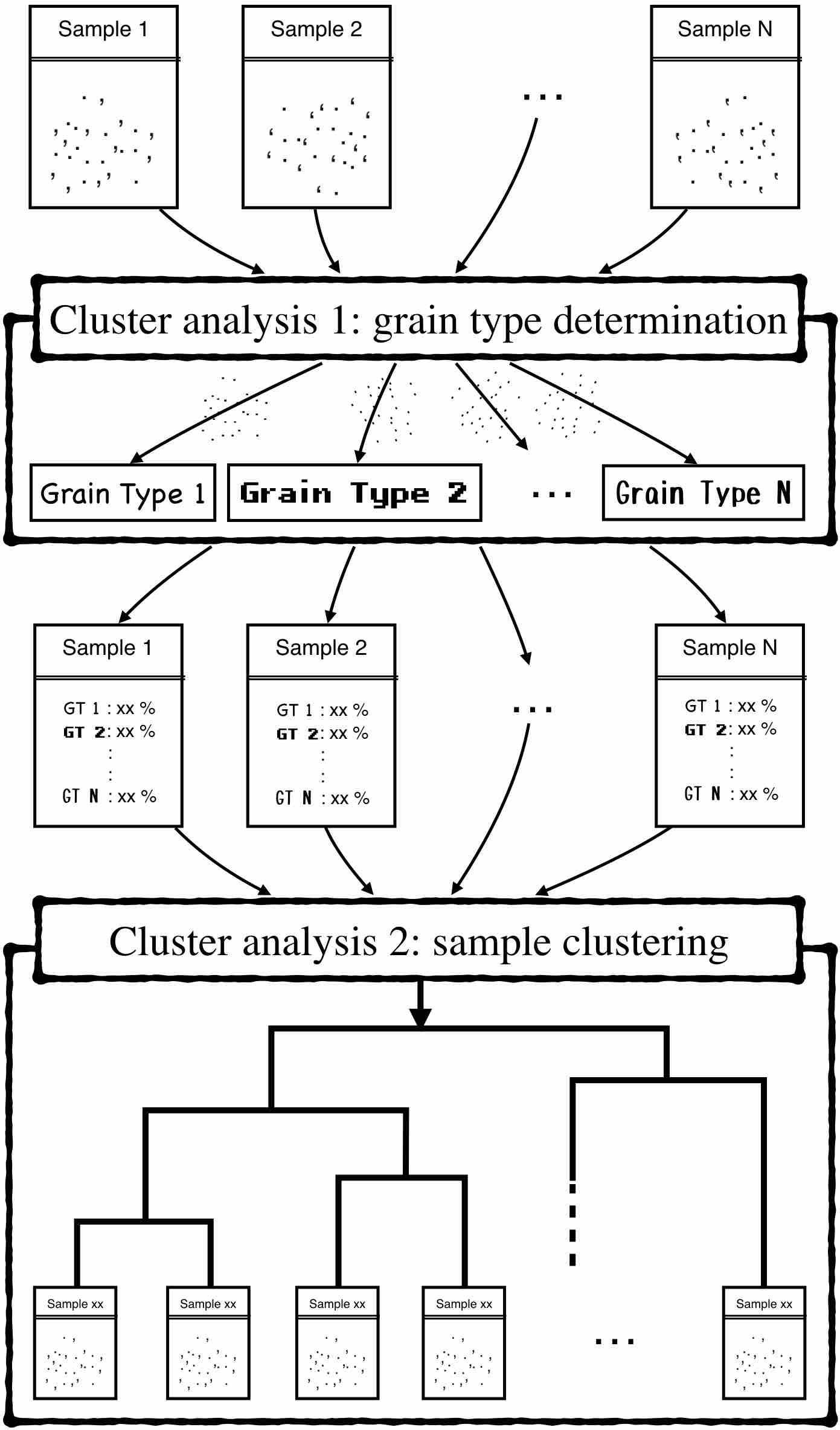}
\caption{Cluster analysis procedure used in this study.}
\label{cluster_procedure.jpg}
\end{center}
\end{figure}

To obtain the feature vector representation, we performed a two-step clustering analysis to 1) categorize whole ash into a small number of grain types, and 2) represent samples by a feature vector composed of the ash ratio (Fig.\ref{cluster_procedure.jpg}).
In the first cluster analysis, we categorized whole ash grains across the entire sample, and then considered each cluster as a statistically determined grain type.
After calculating the grain number percentages of each grain type for all samples, considering the proportions of  grain types as the feature vector for the sample, we then categorized the samples.
We performed two sets of analyses, either including or excluding transparency values (i.e., $I_m$ and $I_{sd}$), to evaluate the effect of transparency on our results.
Because the range of values differ between shape parameters (0--1) and intensity values (0--255), ash grain data were standardized (using the {\tt scale} function, with a mean 0 and a standard deviation 1) before the analysis.
Whole ash grain data include images are shown in the Supplemental Information.

To define grain types, we determined the number of clusters of whole ash grains in the samples.
There are several methods for determining the number of clusters; for example, the R package {\tt NbClust} provides 30 different indices for determining the number of clusters (\cite{Charrad2014}).
In the case of the {\tt NbClust} package, the best number of clusters is determined by majority vote of the optimal numbers of clusters which are defined for each index based on maximum/minimum differentiation (see \cite{Charrad2014} for details).
Some of these indices require a heavy computational burden, particularly considering that our data includes 13,120 ash grains.
In this case, it is recommended by the package authors to use only 18 of the 30 indices, which is more computationally efficient and do not depend on visual inspection.
Therefore using {\tt NbClust} package, we determined the appropriate cluster number for 13,120 ash grains.
We analyzed both cases of including or excluding transparency values.
See \cite{Charrad2014} for details of the indices used in this study.

\section{Results}

We found that there is significant difference in the sample clustering results depending on whether transparency values are used or not.
In the case of using transparency values, we got a consistent result of the sample clustering with their origin.

\subsection{Number of grain types}
\begin{figure}[htbp]
\begin{center}
\includegraphics[width=15cm]{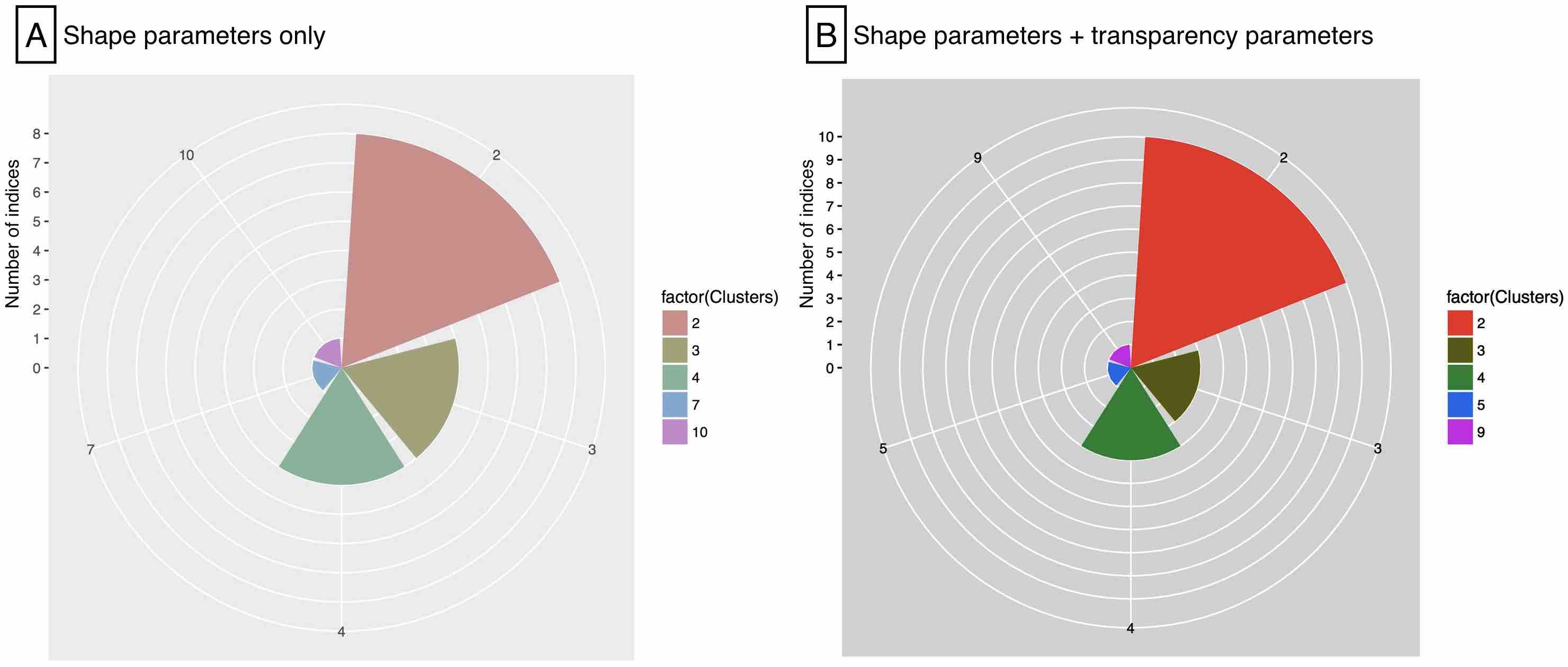}
\caption{The appropriate number of grain type by the {\tt NbClust} analyses.  A: in the case of using grain shape parameters only.  B: in the case of using grain shape parameters and transparency values.  The radius of the pie chart shows the number of indices.  Each color correlate with the number of clusters.}
\label{NbClust_all.jpg}
\end{center}
\end{figure}

According to results of the {\tt NbClust} analysis, the appropriate number of grain type was determined as two (Fig.\ref{NbClust_all.jpg}) for our dataset.
Regardless of whether transparency values are used, the majority of indices showed their best number of clusters ({\tt Best.nc} values in the {\tt NbClust} package) as two.
We checked and compared the results with the expectation of finding a larger number of grain type, but the results were almost equal.
Therefore, the following analyses used two as the number of grain type.

\subsection{Cluster analysis 1: grain type determination}

\subsubsection{I: grain shape only (without transparency values)}\label{CA1_wot}

\begin{figure}[htbp]
\begin{center}
\includegraphics[width=15cm]{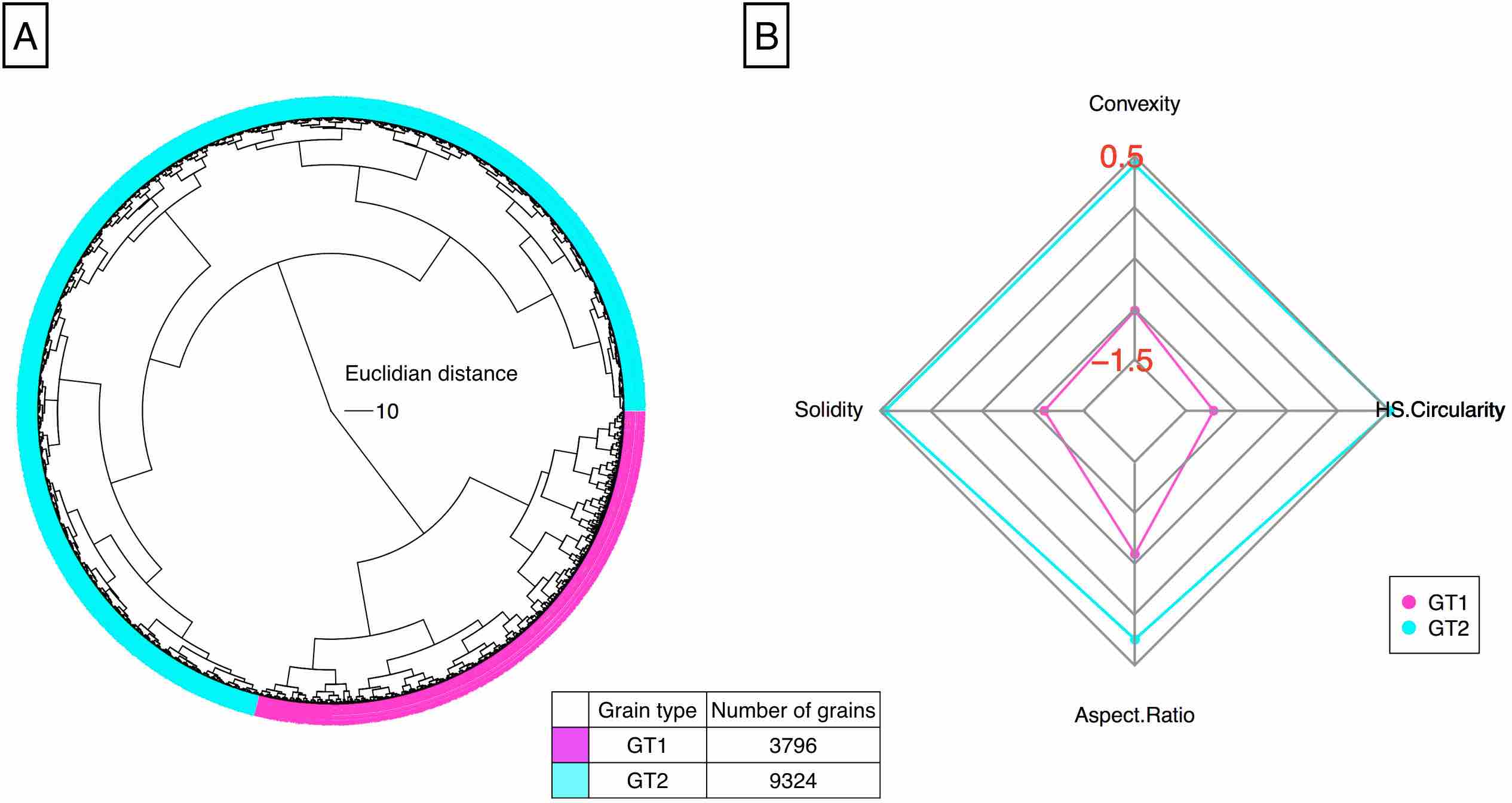}
\caption{Results of cluster analysis 1: grain type determination.  A : dendrogram for each grain type, and D: radar chart for each grain type.}
\label{CA1_wot.jpg}
\end{center}
\end{figure}

\begin{figure}[htbp]
\begin{center}
\includegraphics[width=15cm]{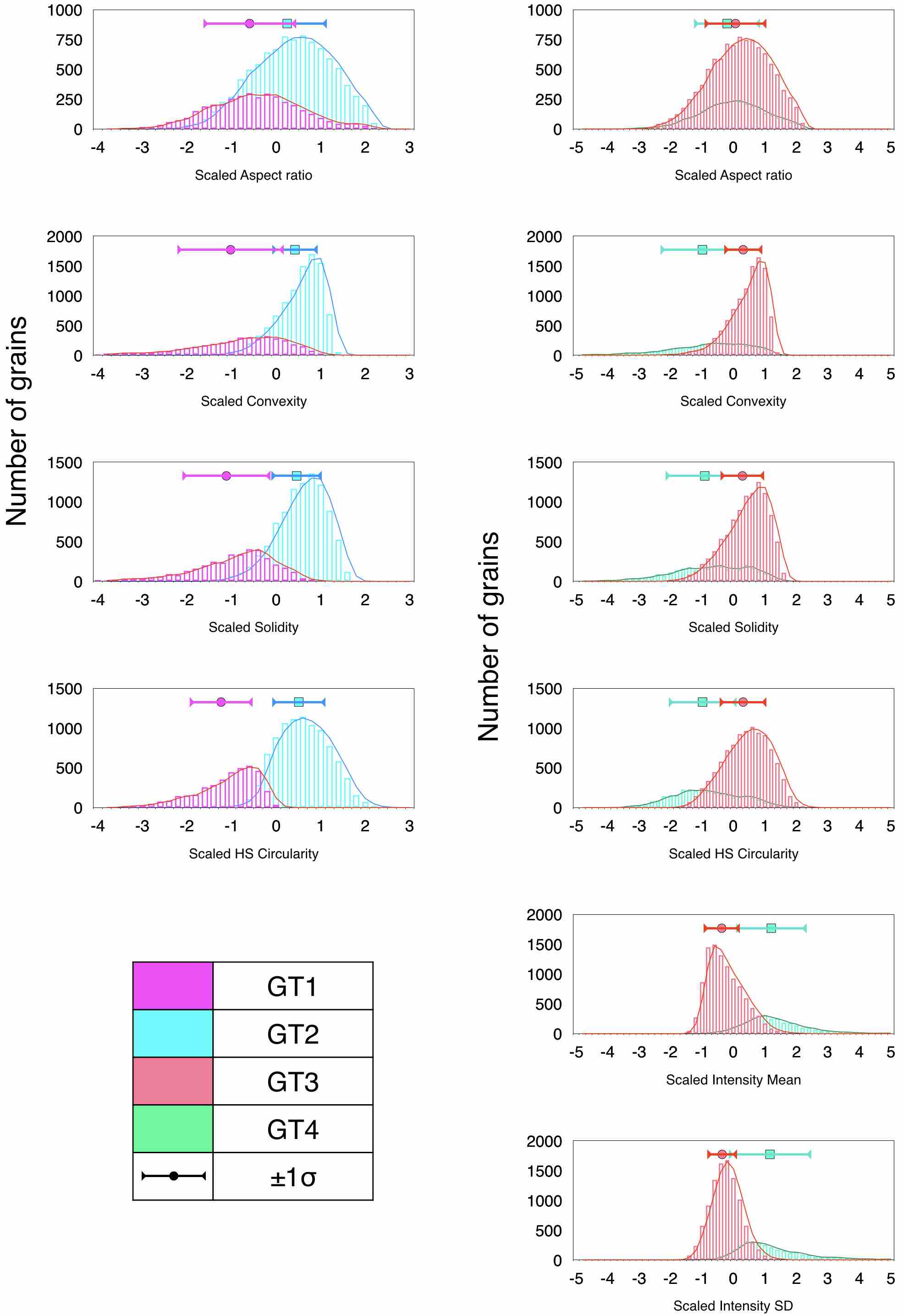}
\caption{Histograms for each scaled parameters in cluster analysis 1.  Left: shape parameters only.  Right: shape parameters and transparency values.}
\label{CA1_histogram.jpg}
\end{center}
\end{figure}

Grains are divided into two clusters: an irregular-shape type (GT1) and a simple-shape type (GT2).
Figs. \ref{CA1_wot.jpg}A and B show the dendrogram and radar chart of centroids for each grain type.
In the grain type determination, the aspect ratio is less effective, and others contribute to the classification (Fig.\ref{CA1_histogram.jpg}).
The larger cluster, GT2, includes 9,324 grains, and the smaller cluster, GT1, has 3,796 grains (Fig. \ref{CA1_wot.jpg}).
GT1 shows low values for all parameters, indicating irregular shape characteristics.
GT2 generally shows high values, indicating rounded and smooth shape characteristics.
Typical images of grains in each grain type are shown in Fig.\ref{Typical_grains.jpg}A.

\subsubsection{II: grain shape and transparency}\label{CA1_wt}

\begin{figure}[htbp]
\begin{center}
\includegraphics[width=15cm]{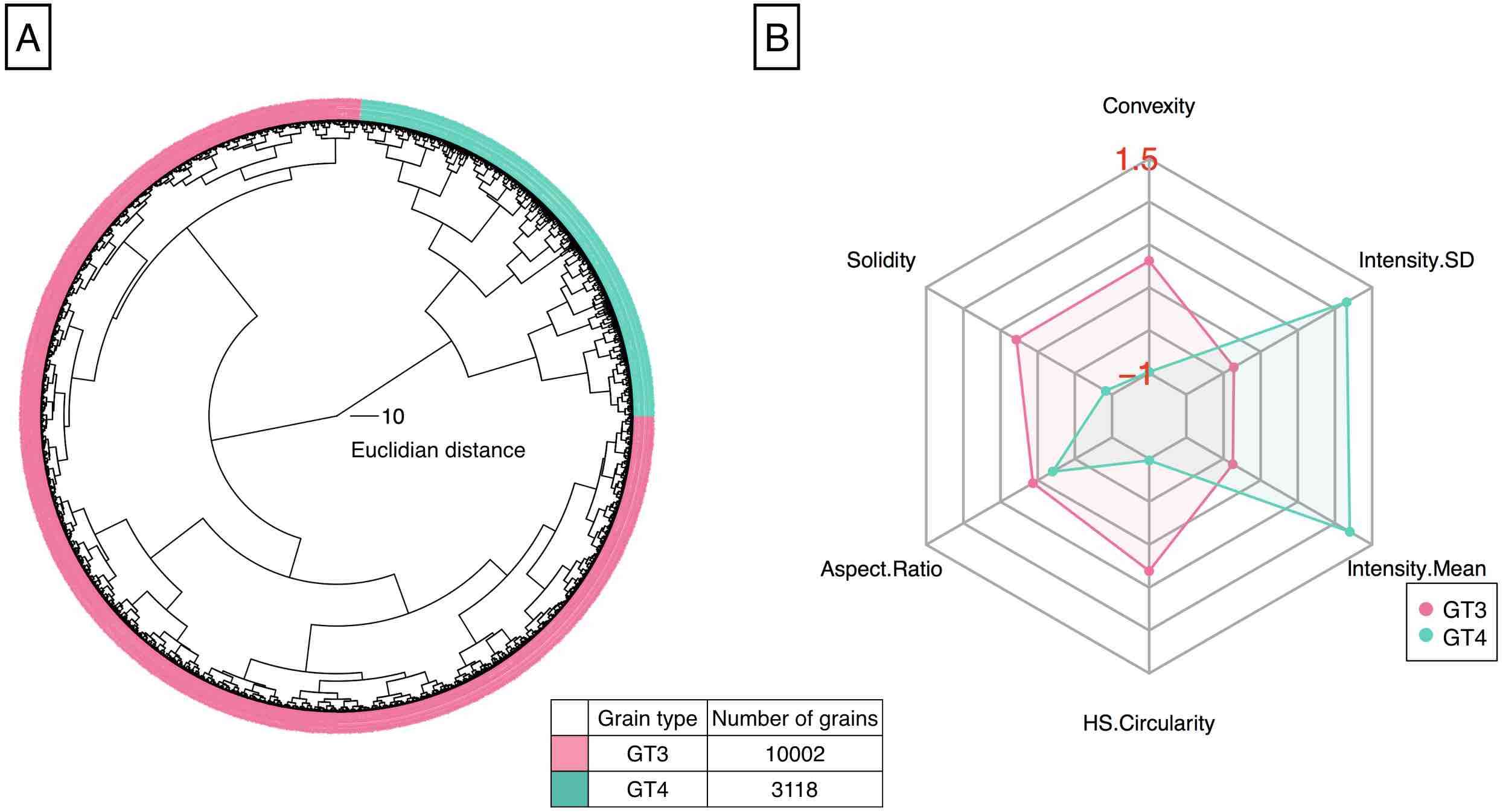}
\caption{Results of cluster analysis 1: grain type determination.  A : dendrogram for each grain type, and D: radar chart for each grain type.}
\label{CA1_wt.jpg}
\end{center}
\end{figure}

\begin{figure}[htbp]
\begin{center}
\includegraphics[width=15cm]{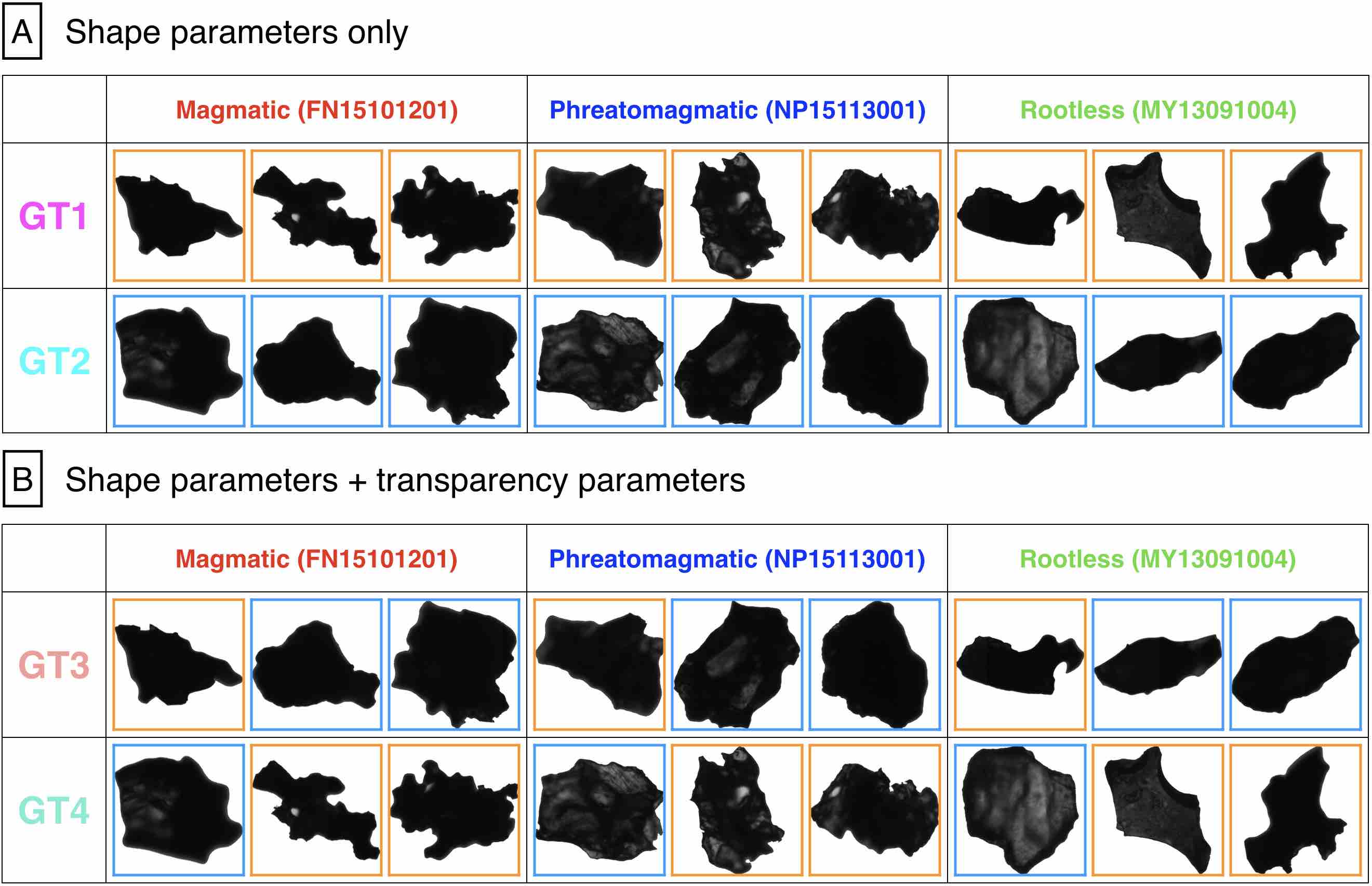}
\caption{Examples of grains in cluster analysis 1 for each grain type.  Orange and blue boxed grains are determined as GT1 and GT2, respectively.}
\label{Typical_grains.jpg}
\end{center}
\end{figure}

Grains are classified into two grain types: a simple-shape opaque type (GT3) and an irregular-shape transparent type (GT4) (Fig. \ref{CA1_wt.jpg}).
As is the case in \ref{CA1_wot}, the aspect ratio does not show significant difference between these two grain types.
GT3 includes 10,002 grains, and has the highest values of the four grain shape parameters, which indicate a relatively rounded and smooth appearance.
The transparency values of GT3 grains are low, indicating its opacity.
GT4, which includes 3,118 grains, has the lowest values of the four grain shape parameters, suggesting irregular grain shapes.
Grains assigned to GT4 are transparent, as suggested by their higher transparency values.
Typical images of grains in each grain type are shown in Fig.\ref{Typical_grains.jpg}B.

\clearpage

\subsection{Cluster analysis 2: sample clustering}

Percentage of numbers of grains in each grain type was calculated for every sample (Table \ref{GT_percentages}).

Components of GT1 grains reached 50 \% only for NP15113001, NP15113002, and NP15113003.
GT2 grains are dominant (over 80 \%) for FN15101206, FN15101208, NP16102707, MY13091004, MY13091306, MY13091402, and MY13092002.

Funabara samples, MY13091006, MY13091306, and MY13091402 has significantly high component of GT3 grains (more than 90 \%).
NP15113001, NP15113002, and NP15113003 has more than 50 \% of components for GT4 grains.

Based on the proportions of grain types, we performed a cluster analysis for 18 samples from three origins.

\newpage
\begin{table}[htbp]
\begin{center}
\caption{Grain type proportions for each sample.}
\begin{tabular}{|l||l|l||l|l|l||}
\hline
Sample ID  & GT1 [\%] & GT2 [\%] & GT3 [\%] & GT4 [\%]\\
 \hline
FN15101201 & 32.82 & 67.18& 93.13 & 6.87 \\
FN15101205 & 26.72 & 73.28 & 95.04 & 4.96 \\
FN15101206 & 14.56 & 85.44 & 97.09 & 2.91 \\
FN15101207 & 28.74 & 71.26 & 97.7 & 2.3 \\
FN15101208 & 11.31 & 88.69 & 100 & 0 \\
\hline
NP15113001 & 50.62 & 49.38& 43.92 & 56.08 \\
NP15113002 & 51.06 & 48.94& 49.79 & 50.21 \\
NP15113003 & 53.04 & 46.96& 50 & 50 \\
NP15113004 & 23.82 & 76.18& 73.33 & 26.67 \\
NP15113005 & 34.32 & 65.68& 68.36 & 31.64 \\
NP15113006 & 25.88 & 74.12& 71.61 & 28.39 \\
NP16102407 & 14.95 & 85.05& 86.67 & 13.33 \\
\hline
MY13091004 & 19.83 & 80.17& 88.84 & 11.16 \\
MY13091006 & 22.16 & 77.84& 93.62 & 6.38 \\
MY13091305 & 38.92 & 61.08 & 67.06 & 32.94 \\
MY13091306 & 16.87 & 83.13 & 92.39 & 7.61 \\
MY13091402 & 18.66 & 81.34 & 95.33 & 4.67 \\
MY13092002 & 16.89 & 83.11& 89.74 & 10.26 \\
\hline
\end{tabular}
\label{GT_percentages}
\end{center}
\end{table}

\subsubsection{I: Proportions of GT1 and GT2 grains (based on \ref{CA1_wot})}

Using parameterized grain shape only results in poor correlation of sample classification with its origins (Fig. \ref{CA2.jpg}A).
For example, FN15101201, NP15113005, and MY13091305 are classified similarly, yet they have different origins.
An exception is samples NP15113001, NP15113002, and NP15113003, which have the same origin (collected in Nippana) and are well-categorized due to their higher components of GT1 grains (more than 50 \%).
MY13091004, MY13091006, and MY13091402 are relatively well-categorized, though with NP15113004 from the other origin.

\subsubsection{II: Proportions of GT3 and GT4 grains (based on \ref{CA1_wt})}

\begin{figure}[htbp]
\begin{center}
\includegraphics[width=15cm]{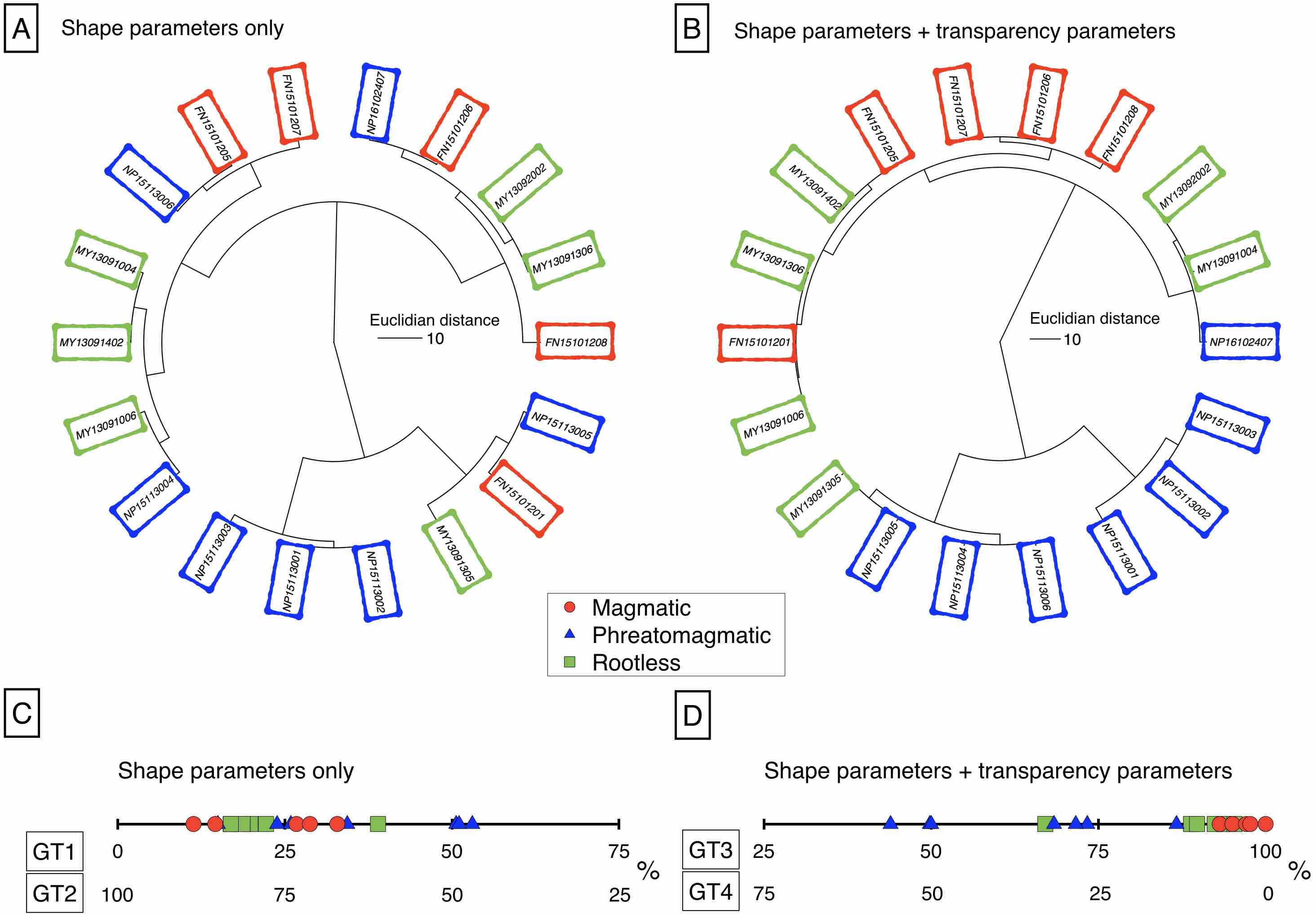}
\caption{Results of cluster analysis 2: sample clustering.  A and C: grain types are determined based on shape parameters only and B and D: grain types are determined based on shape parameters and transparency values. Red, blue, and green boxes indicate magmatic, phreatomagmatic, and rootless samples, respectively.  C and D: number lines for grain type percentages.}
\label{CA2.jpg}
\end{center}
\end{figure}

Samples from Funabara scoria cone and Nippana tuff ring are clearly distinguished when considering both grain shape and transparency (Fig. \ref{CA2.jpg}B).
The dominant component of the Funabara samples is GT3 grains (more than 90 \%, Fig. \ref{CA2.jpg}D, Table \ref{GT_percentages}).
Especially, FN15101206, FN15101207, and FN15101208, upper layers of Funabara samples, are categorized similarly due to their significantly high GT3 grain components.
For samples from Nippana, the stratigraphically lower samples (NP15113001, NP15113002, and NP15113003) are split equally between GT3 and GT4, whereas the stratigraphically upper samples (NP15113004, NP15113005, NP15113006, and NP16102407) are richer in GT3 grains (over 68 \%).
In the dendrogram, NP16102407 was distinguished from other Nippana samples because its component of GT4 grains is low (13.3 \%).
Myvatn samples are dominant in GT3 grains (over 65 \%), however they were not categorized together.
Notably MY13091305 was categorized within Nippana sample dominant cluster due to its higher component of GT4 grains (32.9 \%).
Although they were collected from same rootless cone respectively, MY13091004 and MY13092002 are distinguished with MY13091006 and MY13091306.

\clearpage

\section{Discussion}

\subsection{Verification of introducing grain types by the cluster analysis}

The components of the grain type in each sample are consistent with microscopic observations by human eyes.
As noted in \ref{Nippana tuff ring}, we found two characteristics of grain shape: inwardly convex shape and rectilinear edge.
This difference appeared in our grain types; inwardly convex shape grains corresponds GT1 and GT4, and rectilinear edge grains correspond GT2 and GT3, respectively.
This difference effects to the sample clustering (Fig.\ref{CA2.jpg}).
In human eyes, transparent grains are dominant in MY13091004 and MY13091305, and opaque grains are dominant in MY13091006 and MY13091402.
Our grain types which consider both of grain shape and transparency are consistent with this qualitative observation; MY13091004 and MY13091305 has higher contents of GT4 grains, and MY13091006 and MY13091402 has significantly higher contents of GT3 grains (Table \ref{GT_percentages}).
However, FN15101206, has higher component of brownish yellow grains in microscopic observations, did not show unique grain type component in our analysis.
Possibly, these brownish yellow grains might be appeared as opaque on Morphologi images.

We have clarified that grain transparency plays an important role in the classification of grain and sample types.
Previous studies have considered only grain shape, ignoring the transparency of grains, with the exception of \cite{Miwa2015}.
The identification of transparent grains is important for detecting juvenile glass fragments; however, in this study, as the recommended number of grain types was two, all transparent grains were analyzed together, i.e., it was impossible to distinguish between glass fragments and transparent free-crystals, and between sideromelane and tachylite.
Glass fragments considered to formed by two quenching processes: air-cooling and magma (lava) external water interaction.
Therefore amount of glass fragments is expected to be larger in phreatomagmatic and rootless eruptions (both of air-cooling and magma (lava) external water interaction) than those of magmatic eruptions (air-cooling only).

Generation of free-crystals is thought to dependent on the degree of fragmentation; stronger explosion generates larger numbers of free-crystals.
Assuming same mode, assemblage, and size of phenocryst in host magma (lava), phreatomagmatic eruptions might be generating larger numbers of free-crystals than the others.
Thus, GT4 in the cluster analysis with transparency would be mixture of glass fragments and transparent free-crystals by mechanisms shown above.
Hence this explains difference of the GT3/GT4 among eruption types.

\begin{figure}[htbp]
\begin{center}
\includegraphics[width=15cm]{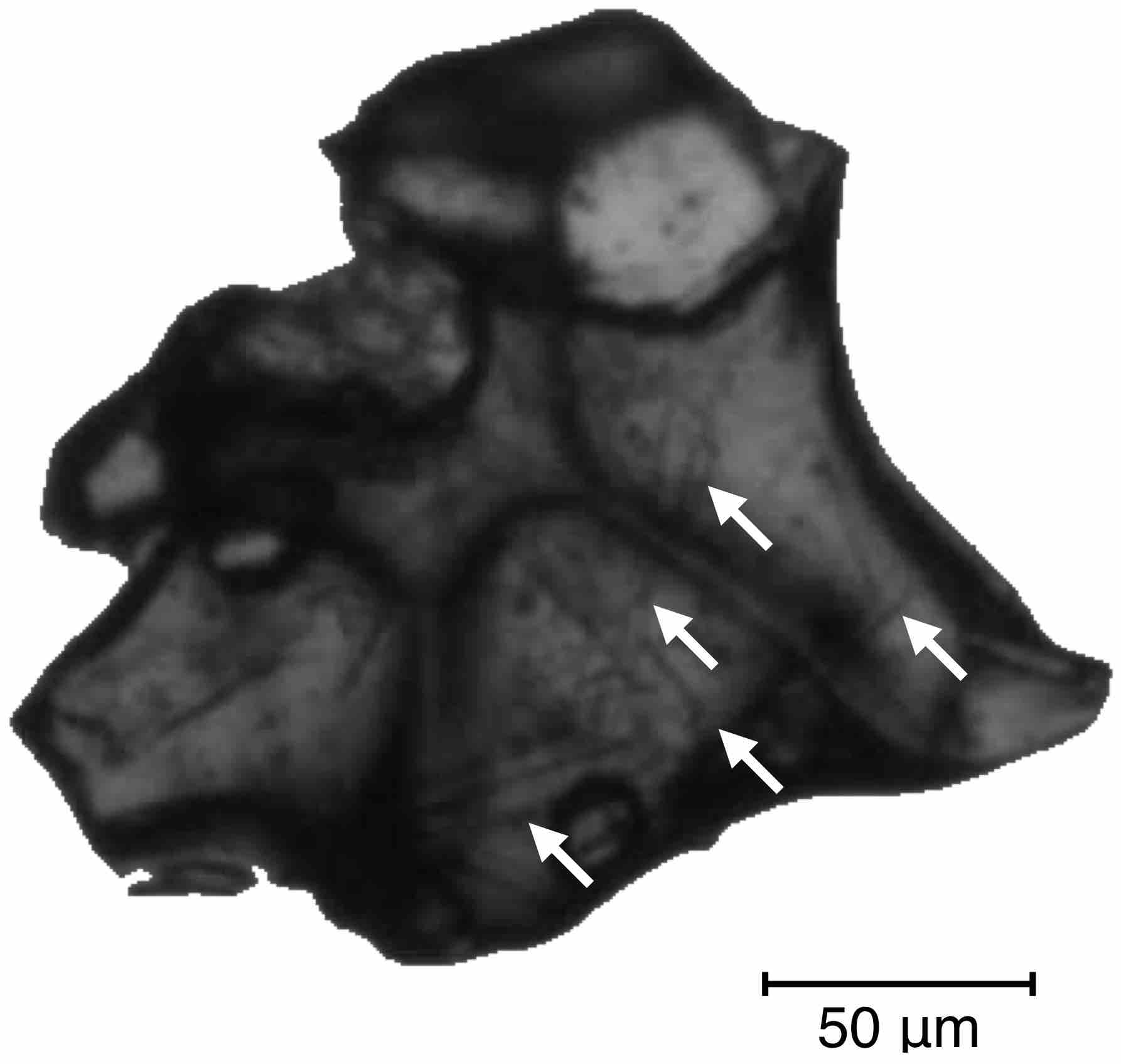}
\caption{An example transparent grain which contains microlites (elongated obstacles, possible plagioclase, white arrows).}
\label{transparent_grain.jpg}
\end{center}
\end{figure}

In the next stage we should explore an efficient criteria to distinguish glass from phenocryst.
For example, quenching of magma (lava) by external water would likely generate a larger amount of sideromelane fragments (e.g., \cite{Taddeucci2004}).
Therefore, identification of sideromelane fragments would be an effective way to discriminate between phreatomagmatic and magmatic samples.
Performing the clustering procedure with a larger number of grain types might enable us to distinguish whether glass fragments are sideromelane or tachylite.
In our ash grain images, it is possible to identify microlites inside of some transparent grains (Fig.\ref{transparent_grain.jpg}).
Microlites in glass grains are thought to decrease transparency of ash grains.
Furthermore, oxidization and existence of microbubbles are also expected to increase opacity of ash grains (e.g., \cite{Yamanoi2008}; \cite{Mujin2014}; \cite{Toramaru1990}).
Hence these controls on the transparency of ash grains should be examined under SEM (Scanning Electron Microscope) observations as a next step.

\subsection{Applicability to identification of explosion/fragmentation process}

By incorporating transparency values into the grain type determination, Funabara (formed by magmatic eruption) and Nippana (formed by phreatomagmatic eruption) samples were clearly distinguished (Fig. \ref{CA2.jpg}A). 
The decisive factor is the amount of GT4 grains; it is less than 7 \% for Funabara samples and up to 56 \% for Nippana samples.
GT4 grains show significant transparency and irregular shape features (Fig. \ref{CA1_wt.jpg}), which might represent quenched glass such as sideromelane.
Therefore, in this clustering scheme, GT4 percentages could be used to infer the magnitude of the effect of external water on volcanic explosions (Fig. \ref{CA2.jpg}D).
A smaller scale cluster structure indicates a possible variation in eruption/fragmentation styles and mechanisms during a volcanic eruption.
For example, Funabara samples were separated consistent with its stratigraphy (lower layers: FN15101201 and FN15101205, upper layers: FN15101206, FN15101207, and FN15101208) (Fig. \ref{CA2.jpg}B).
Lower--middle layers of Nippana samples were also separated consistent with its stratigraphy (lower layers: NP15113001, NP15113002 and NP15113003, middle layers: NP15113004, NP15113005, and NP15113006).
NP16102407 is collected from an upper layer, and distinguished from lower samples.
However, samples from magmatic and rootless eruptions were not completely separated from one another, and one rootless sample (MY13091305) was categorized with the phreatomagmatic samples.
This means the boundary between rootless eruption and phreatomagmatic eruption is obscure.
The highly variable clustering of rootless samples indicates strong variations in the explosion style (e.g., \cite{Fagents2007}; \cite{White2016}).

Our sample classification may be able to extract and distinguish explosion (fragmentation) processes.
It can be possible to simplified explosion (fragmentation) process into two; the vaporization of volatiles originally contained in magma, and the magma (lava)-water interaction analogized as the molten fuel-coolant interaction (MFCI) (e.g., \cite{Dellino2001}).
For example, cluster of NP15113001, NP15113002 and NP15113003 may be represent explosive phase of phreatomagmatic explosions which are driven by both of volatile vaporizations and MFCIs.
In contrast, cluster of NP15113004, NP15113005, and NP15113006 may show subsequent weaken phase of phreatomagmatic explosions which are driven by MFCIs dominantly (volatiles have been degassed, possibly).
This explosion process probably significant in generations of NP16102707, MY13091004, and MY13092002.
FN15101206, FN15101207, and FN15101208 might be formed by mostly pure volatile vaporization-driven explosions.
Though we could not distinguish the other samples which not shown above, these samples might were formed by subequal degree of fragmentation in fine balances of volatile vaporizations and MFCIs (and perhaps other processes), which may be able to distinguish in further analyses.
Thus, our classification method will be contribute to extract explosion/fragmentation processes from volcanic ash grain data.

\subsection{Development of this procedure in volcanology and other fields}

In this study, we show that using grain transparency as well as grain shape significantly improves the result of the cluster analysis for volcanic ash grains.
Our two-step cluster analysis enabled the comparison of many volcanological samples consisting of thousands of ash grains.
The merit of our procedure is that it can be used for volcanic ash analysis of different outcrops, volcanoes, and eruption styles.
However, to apply this procedure to volcanic ash from non-basaltic and non-monogenetic volcanoes, the effects of different magma compositions and alteration degrees should first be verified.

The estimation of an appropriate cluster number depends on each clustering method.
In this study, we adopted the Ward method because of its popularity.
It is our important future work to use other clustering algorithms developed in machine learning community (\cite{Gokcay2002}; \cite{Hino2014}) to obtain detailed characteristics of grain types.

Furthermore, machine learning techniques would be required to apply the results of our analyses to other ash samples.
Involving our classified volcanic ash images as training data in the supervised learning, machine can identify grains and samples.
For this application, further grain parameters (e.g., RGB (red, green, and blue) color under the incident lighting which have information about oxidization and alteration) and further types of samples (e.g., collected from non-monogenetic, non-basaltic volcanoes).

Although our analyses show the importance of transparency in the clustering of volcanic ash grains, it is not clear whether the average ($I_m$) and standard deviation ($I_{sd}$) of gray scale values of one grain are sufficient for accurate identification.
For instance, \cite{Liu2015} showed that the choice of parameters plays a significant role in statistical analysis.
Despite parameter limitations, the machine learning technique is expected to be useful for direct volcanic ash image interpretation (\cite{Shoji2017}).

The development of our procedure will be useful outside volcanology, such as sedimentrolgy and planetary science (e.g., sample return mission such as for Mars and asteroids).

\section{Conclusions}

We constructed a statistical procedure for comparing volcanic ash samples from several outcrops and volcanoes.
Using the cluster analysis, we set "grain types", then categorized samples by its components of each grain type.
Testing our data set, we found that grain types should be two, and the transparency values of grains is effective to categorize samples consistent with their origins.
This procedure detailed in this study could be used to interpret changes in eruption/fragmentation style during a volcanic event.
Through further analyses and using other machine learning techniques, our procedure can contributes outside volcanology and Earth.

\section*{Acknowledgements}
Grain data in this study were collected using Morphologi G3S at the Geological Survey of Japan, AIST.
Discussion with Hiroaki Sato and Daigo Shoji was helpful to considering transparency of ash grains.
We would like to acknowledge Yusuke Suzuki for help with the field work in the Izu Peninsula, the Tachiiwa corporation, a quarrying company of the Funabara scoria cone, for enabling sampling, Kar\'oly N\'emeth for field work assistance in Miyakejima, and \'Armann H\"oskuldsson, \'Arni Einarsson, Eirik Gjarlow, \'Arni Fridriksson, and Tomotaka Saruya, who assisted R.N. and K.K. in the field work in Myvatn.
The field work in Izu Peninsula and Iceland was funded by the Izu Peninsula Geopark Promotion Council and the Sasakawa Scientific Research Grant from The Japan Science Society for R.N. (25-602), respectively.
H.H. is supported by KAKENHI No.16H02842, and JST CREST  ACA20935.
This study was supported by the Joint Usage/ Research Center program No. 2015-B-04 from the Earthquake Research Institute, the University of Tokyo, and KEKENHI No.17H02063.

\section*{Appendix: in the case of three grain types}

\begin{figure}[htbp]
\begin{center}
\includegraphics[width=15cm]{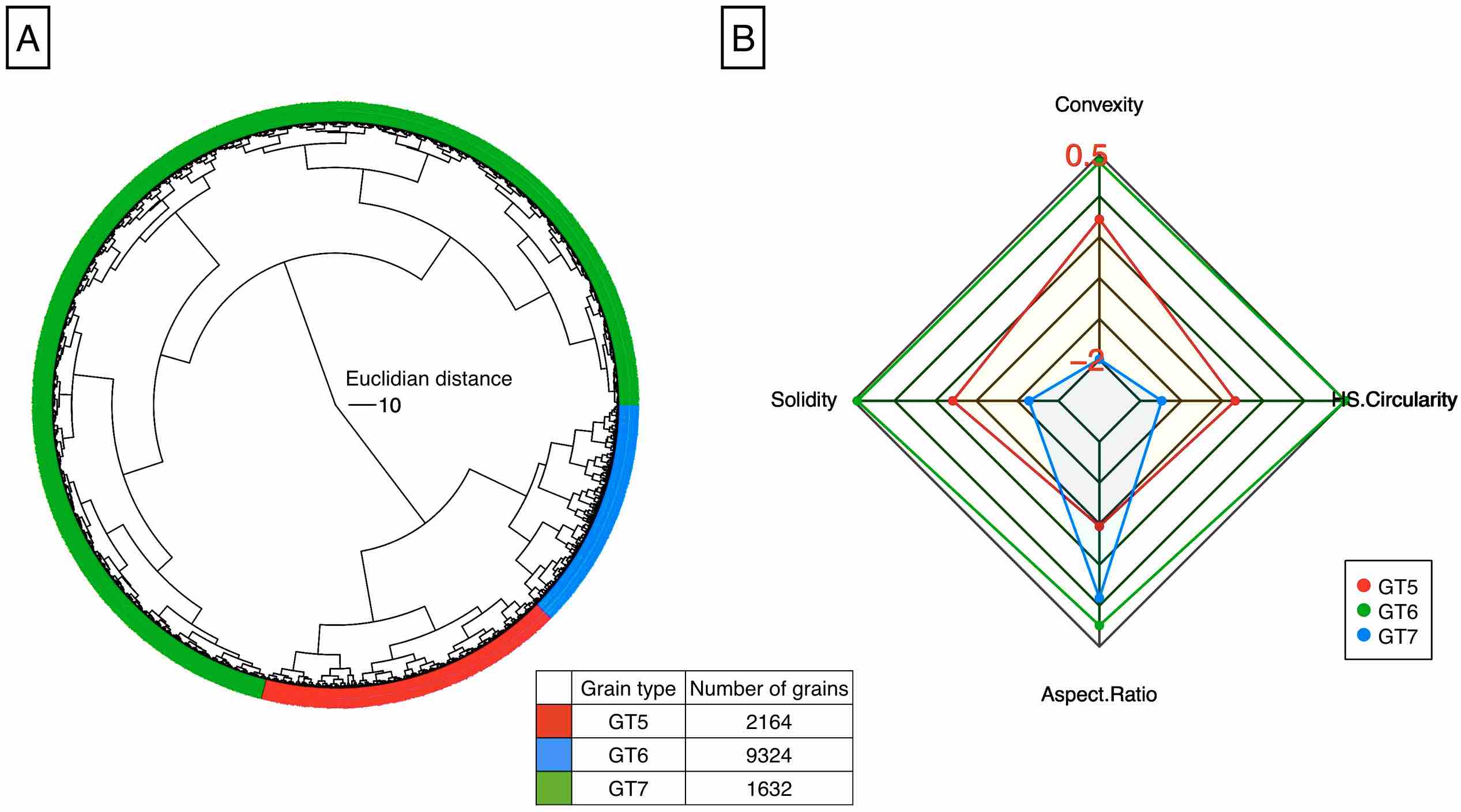}
\caption{Results of cluster analysis 1: grain type determination.  A : dendrogram for each grain type, and D: radar chart for each grain type.}
\label{CA1_wot_GT3.jpg}
\end{center}
\end{figure}

\begin{figure}[htbp]
\begin{center}
\includegraphics[width=15cm]{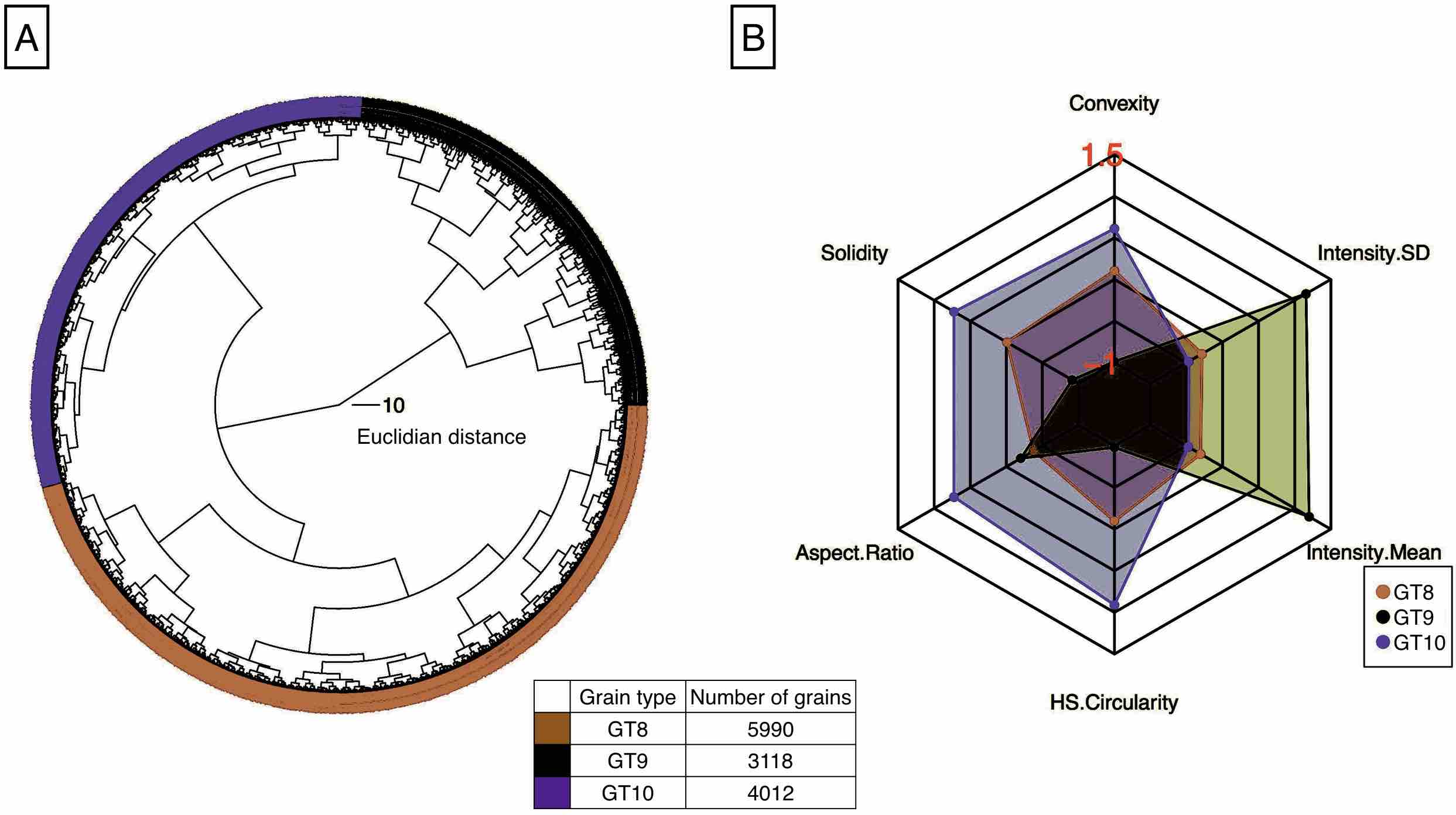}
\caption{Results of cluster analysis 1: grain type determination.  A : dendrogram for each grain type, and D: radar chart for each grain type.}
\label{CA1_wt_GT3.jpg}
\end{center}
\end{figure}

\newpage
\begin{table}[htbp]
\begin{center}
\caption{Grain type percentages for each sample in the case of grain types are three.}
\begin{tabular}{|l||l|l|l||l|l|l|}
\hline
Sample ID  & GT5 [\%] & GT6 [\%] & GT7 [\%] & GT8 [\%] & GT9 [\%] & GT10 [\%]\\
 \hline
FN15101201 & 22.14 & 67.18 & 10.69 & 67.18 & 6.87 & 25.95\\
FN15101205 & 17.18 & 73.28 & 9.54 & 61.83 & 4.96 & 33.21\\
FN15101206 & 8.74 & 85.44 & 5.83 & 57.28 & 2.91 & 39.81\\
FN15101207 & 22.99 & 71.26 & 5.75 & 71.26 & 2.30 & 26.44\\
FN15101208 & 11.31 & 88.69 & 0.00 & 50.00 & 0.00 & 50.00\\
\hline
NP15113001 & 21.34 & 49.38 & 29.28 & 31.23 & 56.08 & 12.70\\
NP15113002 & 23.90 & 48.94 & 27.16 & 36.92 & 50.21 & 12.87\\
NP15113003 & 28.97 & 46.96 & 24.07 & 41.59 & 50.00 & 8.41\\
NP15113004 & 15.91 & 76.18 & 7.91 & 44.71 & 26.67 & 28.62\\
NP15113005 & 18.93 & 65.68 & 15.40 & 45.90 & 31.64 & 22.46\\
NP15113006 & 16.58 & 74.12 & 9.30 & 45.98 & 28.39 & 25.63\\
NP16102407 & 11.12 & 85.05 & 3.82 & 44.96 & 13.33 & 41.71\\
\hline
MY13091004 & 13.87 & 80.17 & 5.96 & 49.08 & 11.16 & 39.76\\
MY13091006 & 15.49 & 77.84 & 6.67 & 55.40 & 6.38 & 38.22\\
MY13091305 & 20.70 & 61.08 & 18.22 & 40.23 & 32.94 & 26.82\\
MY13091306 & 11.94 & 83.13 & 4.93 & 48.51 & 7.61 & 43.88\\
MY13091402 & 11.83 & 81.34 & 6.83 & 50.51 & 4.67 & 44.83\\
MY13092002 & 11.81 & 83.11 & 5.08 & 50.36 & 10.26 & 39.38\\
\hline
\end{tabular}
\label{GT_percentages_GT3}
\end{center}
\end{table}

\begin{figure}[htbp]
\begin{center}
\includegraphics[width=15cm]{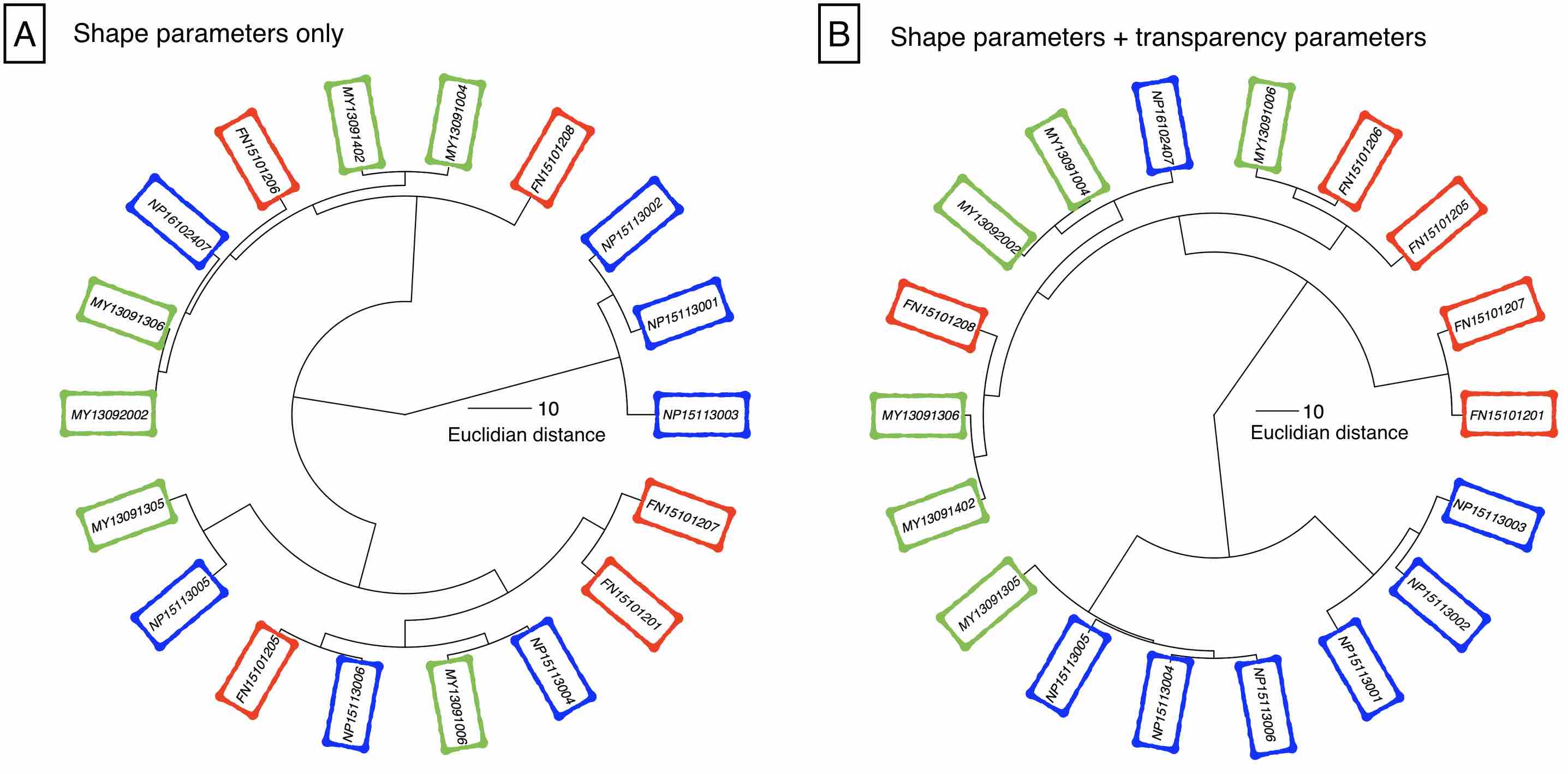}
\caption{Results of cluster analysis 2: sample clustering for in the case of three grain types.  A: grain types are determined based on shape parameters only, and B: grain types are determined based on shape parameters and transparency values. Red, blue, and green boxes indicate magmatic, phreatomagmatic, and rootless samples, respectively.}
\label{CA2_GT3.jpg}
\end{center}
\end{figure}

\beginsupplement

\begin{figure}[htbp]
\begin{center}
\includegraphics[width=15cm]{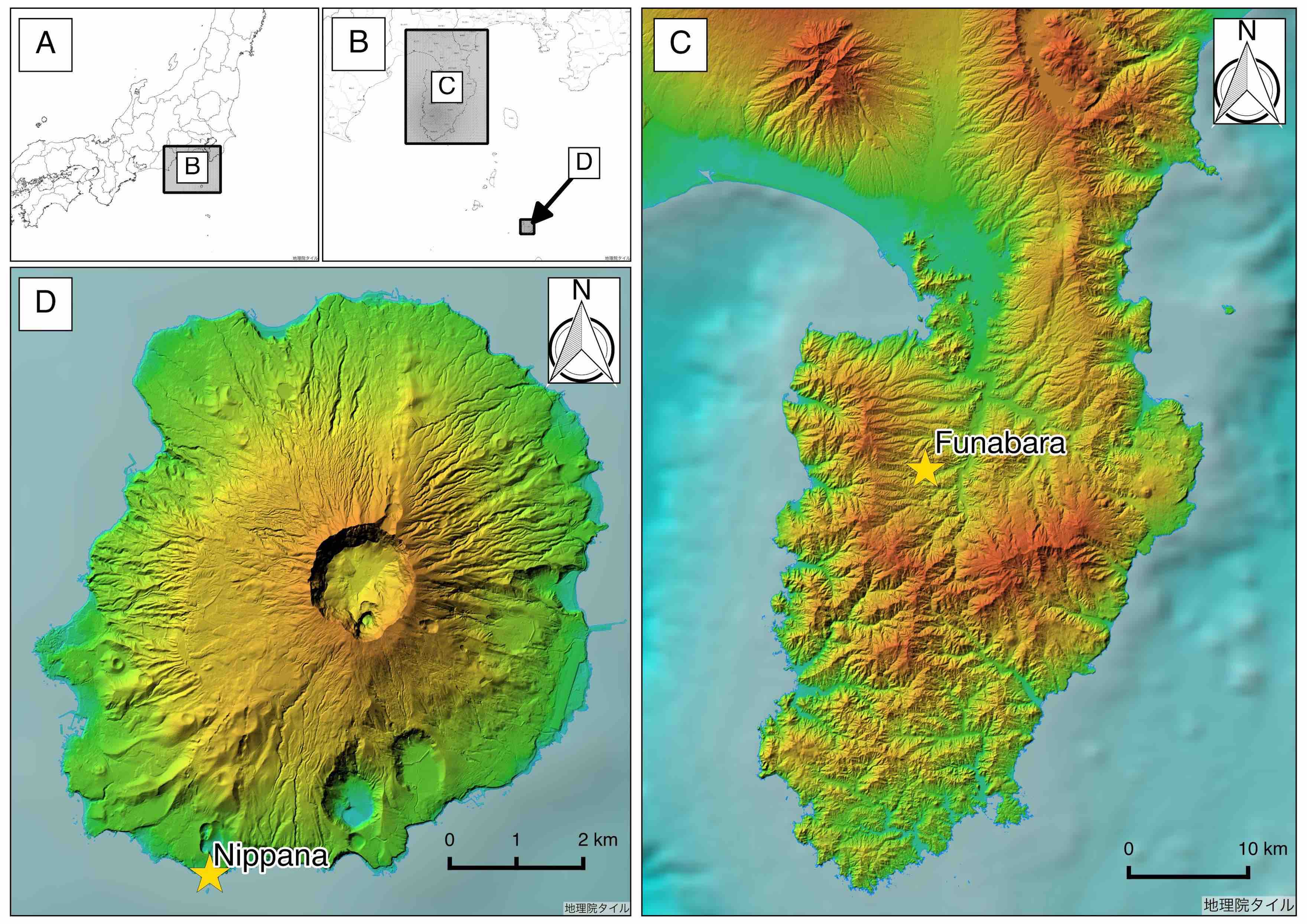}
\caption{Sampling locations for Funabara scoria cone and Nippana tuff ring.  This map is based on "Chiriin Tile" (http://maps.gsi.go.jp) of Geospatial Information Authority of Japan and Hydrographic and Oceanographic Department, Japan Coast Guard.}
\label{Sample_Japan.jpg}
\end{center}
\end{figure}

\begin{figure}[htbp]
\begin{center}
\includegraphics[width=10cm]{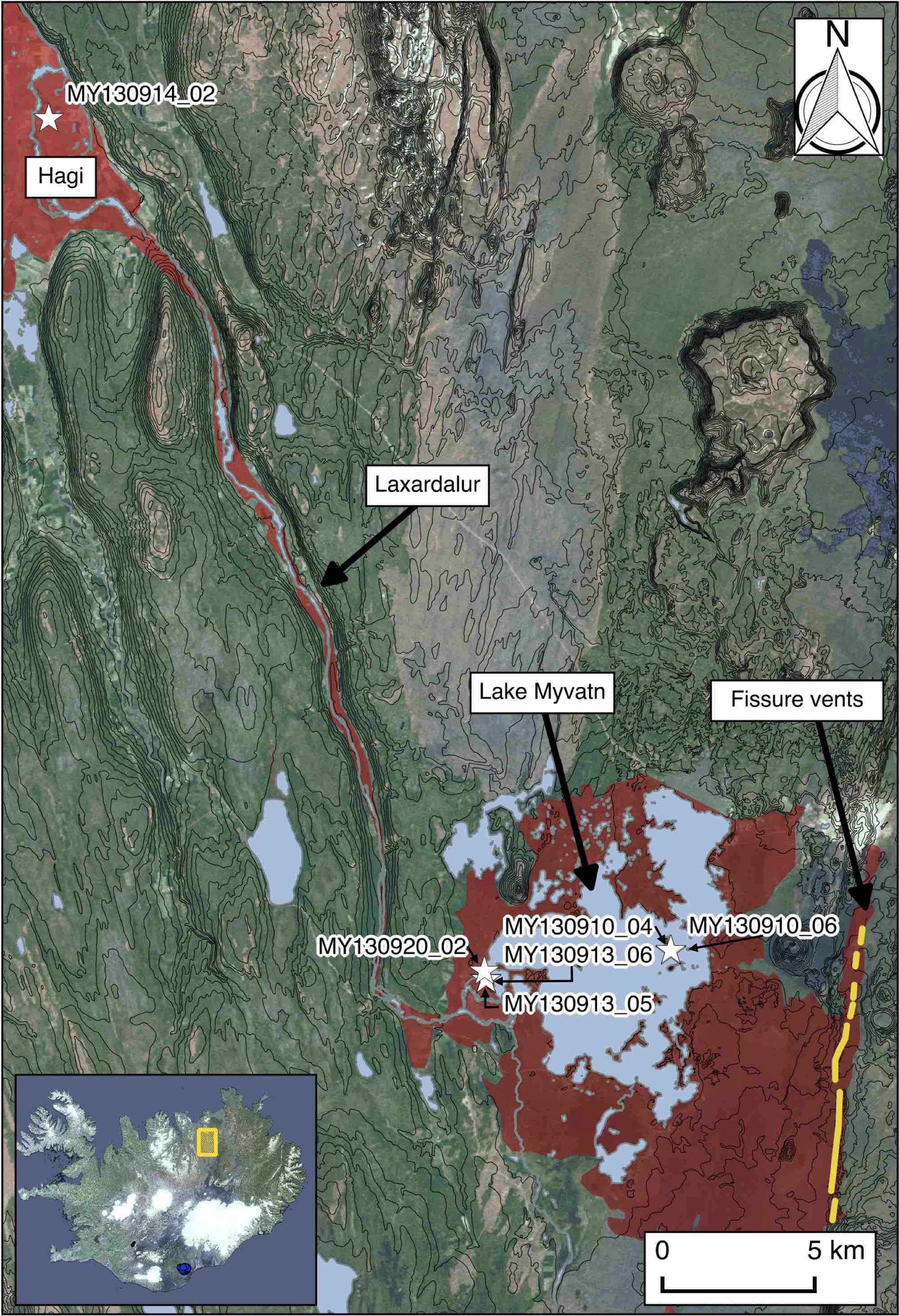}
\caption{Sampling locations for rootless cones in Myvatn, Iceland.  Red region shows extent of the Younger Lax\'a lava.  Yellow lines show fissure vents.  Black tiny lines show 10 m interval topographic contour lines based on the elevation model of Landm\ae lingar \'Islands (LMI).  Background image is the Landsat image mosaic in natural colors (B,G,R), 30 m resolution.  This map is based on data from National Land Survey of Iceland (NLSI).}
\label{Sample_Iceland.jpg}
\end{center}
\end{figure}

\begin{figure}[htbp]
\begin{center}
\includegraphics[width=15cm]{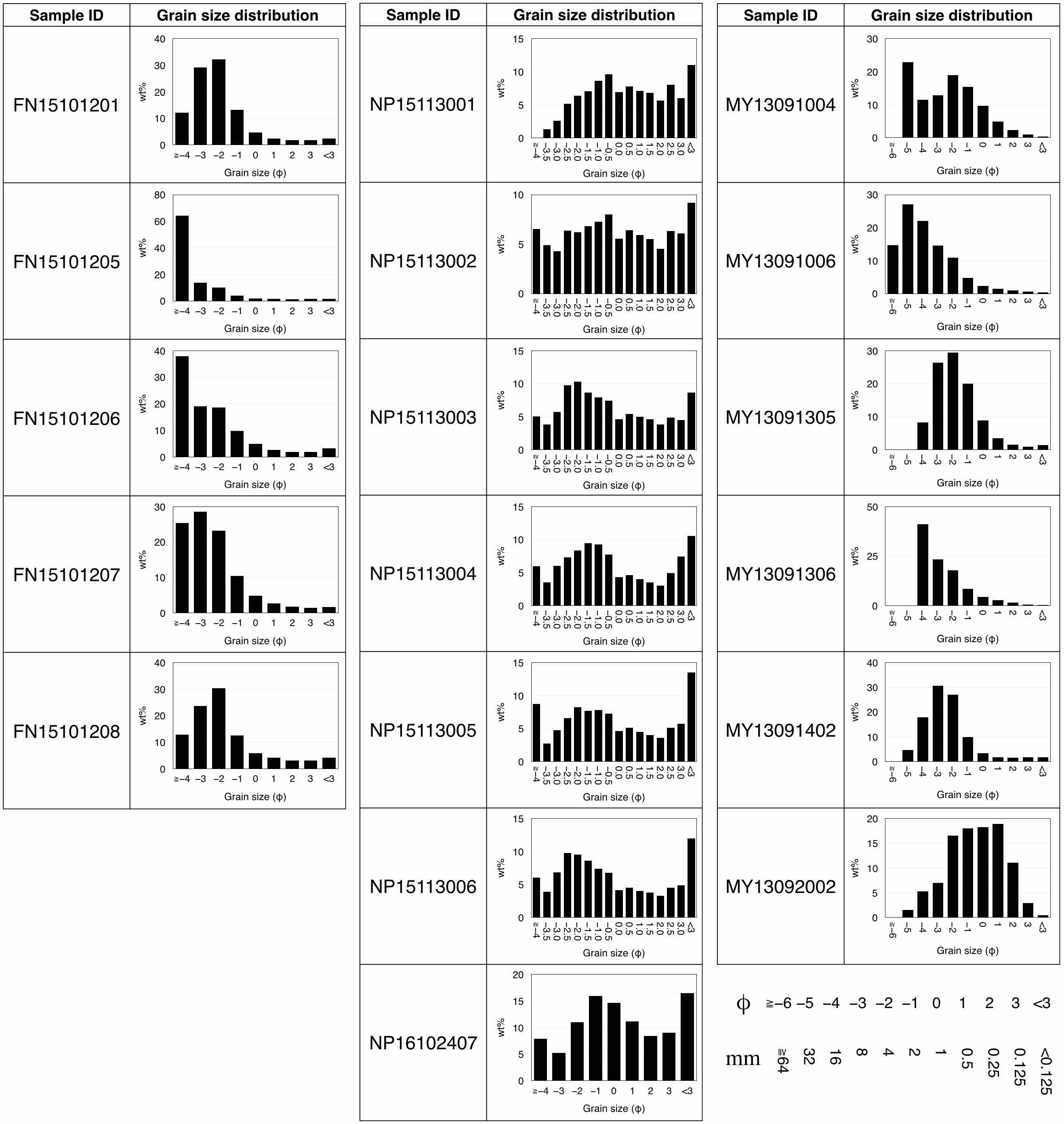}
\caption{The grain size distribution of each sample.}
\label{GSD.jpg}
\end{center}
\end{figure}

\begin{figure}[htbp]
\begin{center}
\includegraphics[width=15cm]{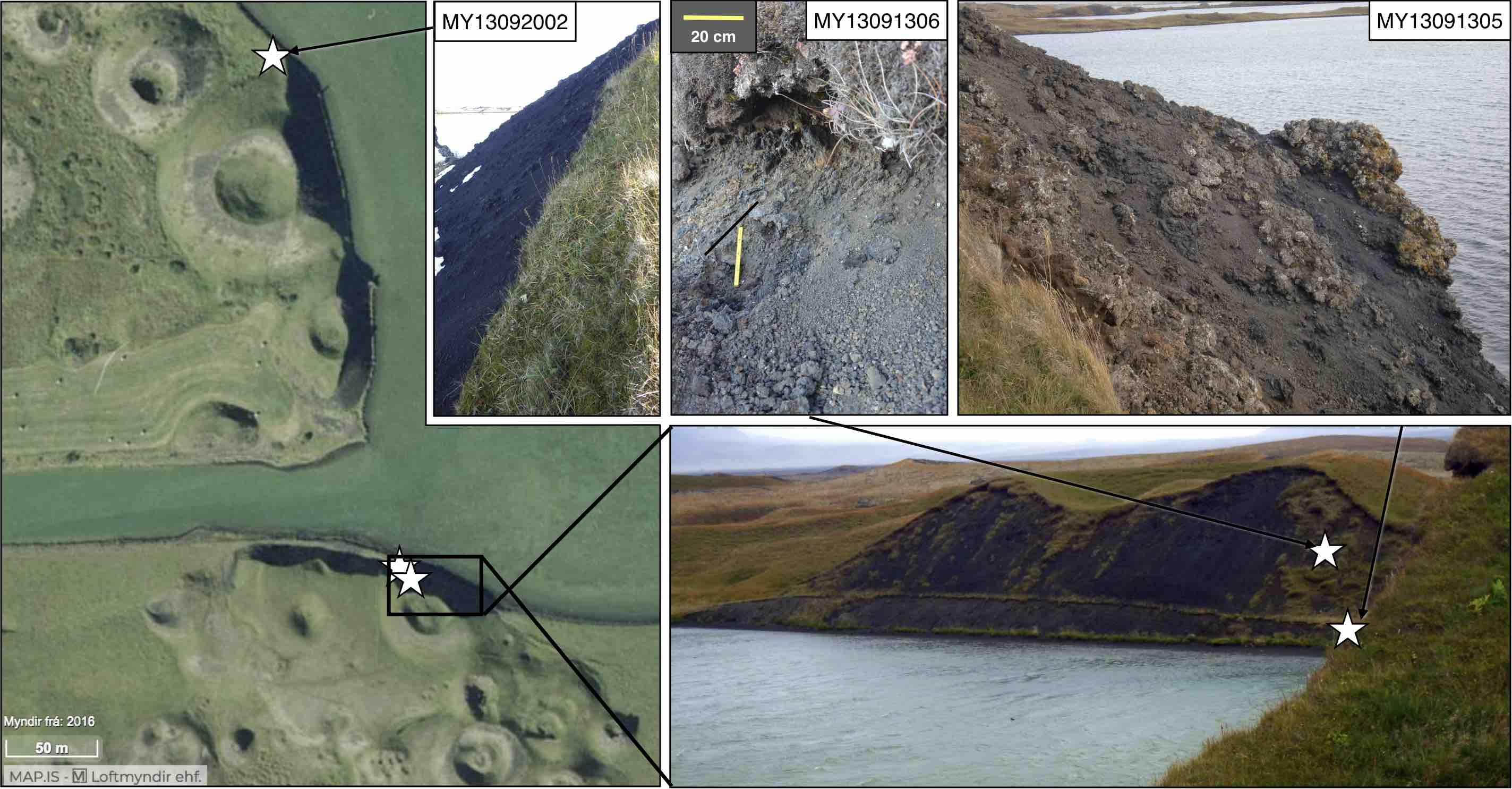}
\caption{Sampling outcrops of Myvatn rootless cones.}
\label{MY130913-20.jpg}
\end{center}
\end{figure}

\begin{figure}[htbp]
\begin{center}
\includegraphics[width=15cm]{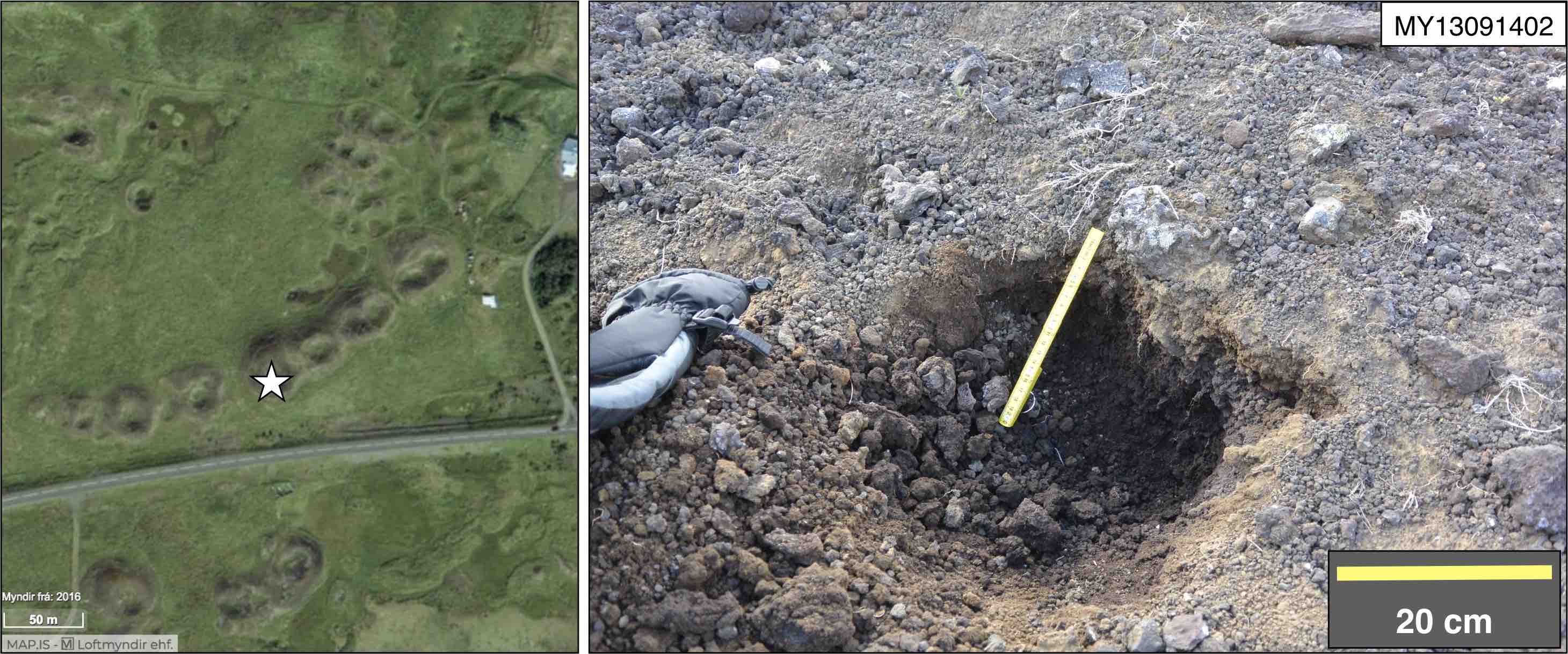}
\caption{Sampling outcrops of Myvatn rootless cones.}
\label{MY13091402.jpg}
\end{center}
\end{figure}

*Title etc. translated by R.N.




\end{document}